\documentclass[draft]{agujournal2019}
\usepackage{url} 
\usepackage{lineno}

\usepackage{soul}
\usepackage{gensymb}

\draftfalse

\journalname{Geophysical Research Letters}

\begin{document}

 \title{SamudrACE: Fast and Accurate Coupled Climate Modeling with 3D Ocean and Atmosphere Emulators}

\authors{James P. C. Duncan\affil{1}, Elynn Wu\affil{1}, Surya Dheeshjith\affil{2}, Adam Subel\affil{2}, Troy Arcomano\affil{1}, Spencer K. Clark\affil{1,4}, Brian Henn\affil{1}, Anna Kwa\affil{1}, Jeremy McGibbon\affil{1}, W. Andre Perkins\affil{1},  William Gregory\affil{3}, Carlos Fernandez-Granda\affil{2, 6}, Julius Busecke\affil{2, 5}, Oliver Watt-Meyer\affil{1}, William J. Hurlin\affil{4}, Alistair Adcroft\affil{3}, Laure Zanna\affil{2,6}, Christopher Bretherton\affil{1}}

\affiliation{1}{Allen Institute for Artificial Intelligence (Ai2), Seattle, USA}
\affiliation{2}{Courant Institute of Mathematical Sciences, New York University, New York, USA}
\affiliation{3}{Princeton University, Princeton, USA}
\affiliation{4}{NOAA/Geophysical Fluid Dynamics Laboratory, Princeton, USA}
\affiliation{5}{Lamont-Doherty Earth Observatory of Columbia University, New York, USA}
\affiliation{6}{Center for Data Science, New York University, New York, USA}

\correspondingauthor{James P. C. Duncan}{jamesd@allenai.org}

\begin{keypoints}
\item SamudrACE is an AI emulator of the GFDL CM4 coupled global climate model (GCM), trained on 150 years of pre-industrial output.
\item Like traditional coupled GCMs, SamudrACE's components were developed independently by distinct teams and coupled via fine-tuning.
\item Running on a single NVIDIA H100 GPU, SamudrACE can simulate 1500 years per day.
\end{keypoints}

\begin{abstract}
Traditional numerical global climate models simulate the full Earth system by exchanging boundary conditions between separate simulators of the atmosphere, ocean, sea ice, land surface, and other geophysical processes. 
This paradigm allows for distributed development of individual components within a common framework, unified by a coupler that handles translation between realms via spatial or temporal alignment and flux exchange.
Following a similar approach adapted for machine learning-based emulators, we present SamudrACE: a coupled global climate model emulator which produces centuries-long simulations at 1-degree horizontal, 6-hourly atmospheric, and 5-daily oceanic resolution, with 145 2D fields spanning 8 atmospheric and 19 oceanic vertical levels, plus sea ice, surface, and top-of-atmosphere variables.
SamudrACE is highly stable and has low climate biases comparable to those of its components with prescribed boundary forcing, with realistic variability in coupled climate phenomena such as ENSO that is not possible to simulate in uncoupled mode.
\end{abstract}

\section*{Plain Language Summary}
Climate scientists use computer models to understand how different parts of the Earth system, like the atmosphere and ocean, work together.
Traditionally, these models are built with separate components that exchange information. 
We applied this same approach to new, much faster models based on artificial intelligence (AI). 
We connected an AI atmosphere model (called ACE) to an AI ocean model (called Samudra) to create a new, coupled AI model called SamudrACE, capable of simulating the full Earth climate evolution.
Our combined model runs stably for centuries, producing accurate, high-quality climate simulations. 
A key success is that by linking the ocean and atmosphere, SamudrACE can realistically simulate complex, large-scale climate phenomena like the El Niño-Southern Oscillation (ENSO), something the individual AI models cannot do on their own. 
This work demonstrates a successful strategy for building fast and powerful AI-based tools to study long-term climate evolution more efficiently.

\section{Introduction}

The advent and success of machine learning (ML)-based weather prediction~\cite{pathak2022fourcastnet,bi2023accurate,lam2023learning} has led to similarly data-driven global atmosphere emulators trained on output of numerical models, such as the atmosphere-only version of the Ai2 Climate Emulator (ACE)~\cite{watt-meyer_ace_2023}.
Since then, atmosphere model emulators have continued to mature and support Atmosphere Model Intercomparison Project (AMIP)~\cite{gates1992ams} compatible simulations~\cite{watt2024ace2,kochkov2024neural,chapman2025camulator}.

This paper will demonstrate early progress toward the natural next step in this progression, a global climate model (GCM) emulator, which consists of modular coupled atmosphere, sea ice, land, and ocean emulators, capable of running the Coupled Model Intercomparison Program (CMIP) DECK simulation suite~\cite{eyring2016overview}.  This could later be extended to incorporate
other components of the Earth system (e.g., biogeochemical processes).

Coupled atmosphere and ocean emulation is needed to learn and generate realistic climate trends (e.g., through the time-evolving spatial patterns of ocean heat uptake and sea-surface temperature rise).  It is also needed to generate the variability in physical phenomena that emerge through the realistic interaction and coupled evolution of atmospheric surface forcing and upper ocean response, such as El Niño-Southern Oscillation (ENSO) variability~\cite{zebiak1987model}.  

Several recent papers have incorporated simplified forms of ocean coupling into ML atmospheric emulators, e.g. by use of a physically-based slab ocean model~\cite{clark2024ace2} expressed in PyTorch~\cite{paszke2019pytorch}, or by prognostically emulating sea surface temperature (SST)~\cite{cresswell2024deep}, or with the addition of near-surface temperature on a limited number of upper-ocean levels~\cite{wang2024coupled}.  While these approaches enable accurate seasonal forecasts~\cite{wang2024coupled} and stable simulation of present-day~\cite{cresswell2024deep} and CO$_2$-modulated ~\cite{clark2024ace2} equilibrium climate, their simplified ocean representations are insufficient to support accurate coupled atmosphere-ocean variability such as ENSO.

To go further toward emulation of climate-coupled phenomena requires resolving the full extent of the ocean in order to simulate ocean circulation and response to atmospheric forcing on annual and decadal timescales.
Fortunately,  three-dimensional ML ocean emulators have been recently developed for data-driven ocean forecasting on timescales up to 1--2 years~\cite{chen2023fuxi,el2024glonet,xiong2023ai,wang2024xihe} and for longer-running simulations forced by specified time-evolving atmospheric conditions ~\cite{dheeshjith2025samudra}.
Specifically, Samudra~\cite{dheeshjith2025samudra} stably emulates GFDL's Ocean Model v4 (OM4) and reproduces ocean dynamics on decadal timescales when forced from above by the net downward heat flux (at the ocean surface, or beneath sea ice where present) and surface wind stresses.
Similarly, successful emulators have been developed for sea ice~\cite{durand2024data}.

However, these advances in component model emulation do not necessarily enable their coupling. So far, there has not been a data-driven approach capable of successful 3D coupled emulation of the full vertical extents of the atmosphere and ocean.
Full Earth system emulation introduces complexities beyond component modeling, including reconciling distinct vertical coordinates, managing new internal boundaries, and selecting between physical or latent space coupling. Crucially, the system must bridge vast spatiotemporal scales -- from fast weather to slow ocean dynamics -- to capture emergent coupled phenomena like ENSO.

In this paper, we present SamudrACE, which is constructed by coupling the ACE2 3D atmosphere emulator to the Samudra 3D ocean emulator which has been extended to predict sea-ice concentration and thickness.  Both components are emulated at 1$\degree$ lat/lon horizontal resolution.  We train SamudrACE to emulate a fully coupled 200-year simulation\footnote{For convenience in the remainder of the text and figures, we use ``CM4" interchangeably to denote both the coupled GCM and the 200-year simulation output from that model.)} by the GFDL CM4 physics-based coupled GCM with constant pre-industrial greenhouse gas and aerosol concentrations and a repeating annual cycle of insolation~\cite{held2019structure}.

\section{Materials and Methods}

\subsection{Component emulators: Samudra and ACE2}

Analogous to CM4's coupling~\cite{balaji2011flexible} of the components AM4/LM4~\cite{zhao2018gfdl,zhao2018gfdl} to OM4/SIS2~\cite{adcroft2019gfdl}, SamudrACE is composed of two independent component emulators and a coupler that handles communication between the two. Specifically, SamudrACE couples the ACE2~\cite{watt2024ace2} atmosphere and land surface emulator to the Samudra~\cite{dheeshjith2025samudra} ocean emulator.
To facilitate this coupling, both ACE2 and Samudra are pretrained from random weight initializations on the CM4 simulation outputs with prescribed forcings, a key step that we call ``uncoupled pretraining" and describe in detail below.
Additionally, we modified Samudra to prognose sea ice variables, including its concentration and mass, which plays a crucial role in SamudrACE's coupler.
After pretraining, ACE2 and Samudra are coupled and fine-tuned in two stages: first only updating Samudra's weights and then jointly updating both components.
SamudrACE is the end result of this pretraining, coupling, and fine-tuning pipeline.
Tables S1 and S2 provide a complete listing of the SamudrACE component outputs.

\subsection{The SamudrACE coupler}

\begin{figure}
\noindent\includegraphics[width=\textwidth]{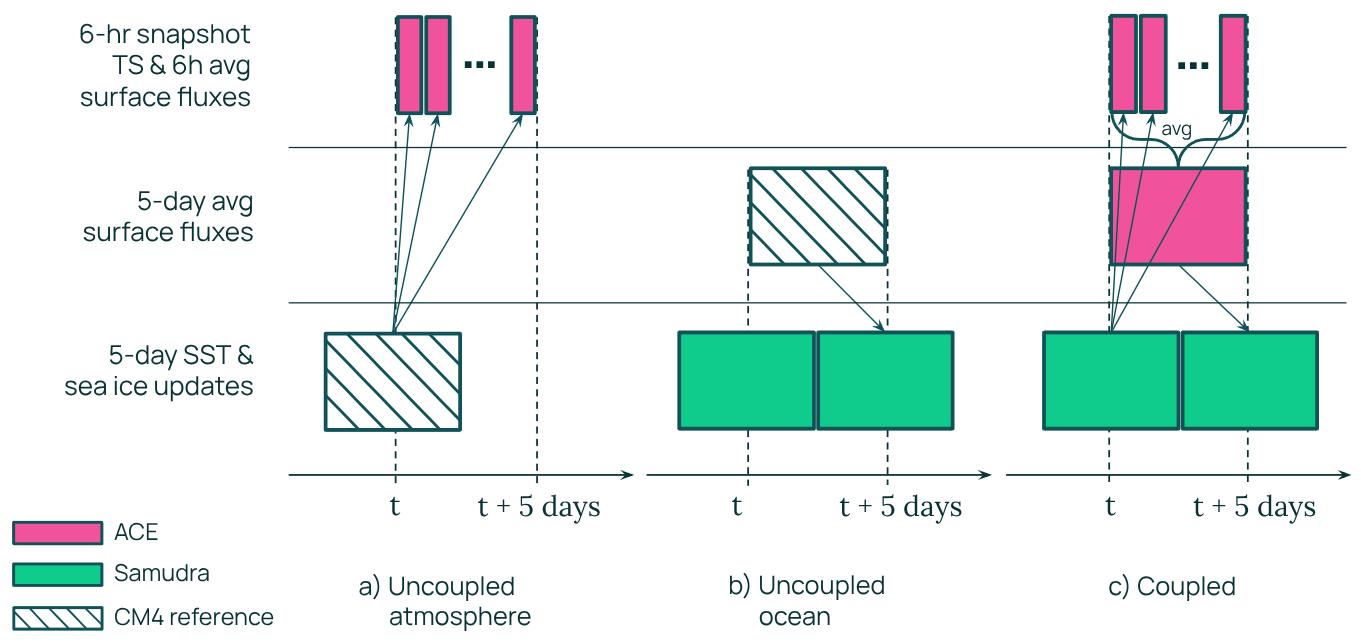}
\caption{
A single 5-day forward step in uncoupled (a, b) and coupled (c) modes.
In uncoupled mode, ACE and Samudra are forced by the appropriate CM4 reference fields.
Uncoupled ACE is forced by the reference 5-day average SST and sea ice concentration, stepping forward 6 hours at a time until reaching 5 days, at which point the next 5-day average forcing is prescribed. 
For uncoupled Samudra, we use the 5-day average of the CM4 reference wind stress, precipitation, and surface fluxes as prescribed forcing inputs.
In coupled mode, aggregation of the diagnostic 6-hour average surface boundary conditions to 5-day averages is done online as ACE completes 20 forward steps.
The generated 5-day average is then passed as input to Samudra, which generates a single 5-day forward step. 
Samudra's generated SST and sea ice states will then be used to force ACE in the next iteration of the coupler loop.
}
\label{fig:coupling}
\end{figure}

Figure \ref{fig:coupling}c provides a schematic of the SamudrACE coupler.
The ACE2 model simulates the atmosphere in 6-hour increments over a 5-day period. 
At the end of each step, core prognostic variables, such as temperature, are saved as an instantaneous snapshot. 
In contrast, diagnostic boundary fluxes are calculated as the average value over that same 6-hour interval, with the averaging period ending at the snapshot time.
For each forward step, ACE2 is forced by the sea ice fraction and SST from time $t$.
Once ACE2 completes 20 forward steps the coupler aggregates the 6-hour average boundary fluxes into a single 5 day average.
The generated ocean surface fluxes of energy, moisture and momentum are then used to force a single step of Samudra, which evolves the ocean state forward in time by 5 days to catch up to the atmosphere. 
The new SST state is then prescribed onto the final generated global surface temperature $T_S$ state in preparation to resume atmosphere forward stepping.

The 6-hour and 5-day timesteps chosen for the atmosphere and ocean emulators reflect their respective timescales of change on the O(100~km) grid scale used in this study.  Using a 20-fold longer ocean time step than atmosphere time step introduces complications in training of the coupled emulator, but also enables simulations that have both realistic variability in both components. 
The SamudrACE coupler is heavily influenced by GFDL's Flexible Modeling System (FMS) coupler~\cite{balaji2011flexible} used in CM4 simulations. 
However, unlike in typical physics-based coupled models, ACE2 internally predicts the surface flux components, rather than these being computed by the ocean or land surface models.

Our choice of coupling procedure is an important physics-informed design element of SamudrACE. The flow of information at component interfaces is column-local and constrained so that the atmosphere forces the ocean with surface fluxes and the ocean forces the atmosphere with SST and sea ice. Enforcing this exchange of information between components in physical state space via a coupling procedure avoids the exchange of nonphysical latent representations between the two models. ACE2 and Samudra were each designed from the start with coupling in mind by learning the diagnostic variables (surface turbulent and radiative fluxes, plus surface precipitation) needed for a physically justifiable coupling procedure and temperature and salinity budget closure.

\subsection{Pretraining ACE2}

We follow ACE2's training protocol \cite{watt2024ace2} with two additional diagnostic variables -- surface zonal and meridional wind stress. 
Training data come from the atmosphere component output of CM4 at 6-hourly temporal resolution, with the exception of surface temperature over ocean, sea ice fraction $f_i$, and ocean fraction $f_o$ (derived from $f_i$). 
For each 5-day window aligned with the ocean data's timesteps, the ocean-covered grid cells of 6-hourly $T_S$ are held fixed at their values at the beginning of the 5-day window, while allowing grid cells with sea ice or land to evolve with the usual 6-hour timestep. 
Similarly, all 20 snapshots of $f_i$ and $f_o$ will be identical to the first snapshot.
We provide ACE2 pretraining details in Text S4 and training losses and inference RMSE in Figure S1b.

\subsection{Pretraining SamudraI}
\label{section:pretrain-ocean}
Our implementation of Samudra~\cite{dheeshjith2025samudra} modifies the original protocol as follows:
\begin{enumerate}
\item
\textbf{Pretraining dataset}: 
We pretrain Samudra on 5-day mean OM4 output.  For consistency, this is taken from the same segment of our reference CM4 simulation as used for ACE2 pretraining.
\item
\textbf{Prognostic Variables and Forcing}: We expanded Samudra to include sea ice concentration (SIC) and thickness, which reduced Arctic upper-layer salinity biases (Figure S21).
We refer to this version as "SamudraI." 
The model is forced by ACE surface conditions (wind stress, heat, and water fluxes) above the sea ice, time-averaged over 5-day blocks.
The original Samudra was forced only by the top-of-ocean heat flux and surface stress, calculated below any sea ice. All forcings are time-averaged over 5-day blocks.
\item \textbf{Time Stepping}: We simplified the architecture to use a single input timestep ($P=1$) and four autoregressive steps ($N=4$) to reduce computational cost without degrading global time-mean bias.
\end{enumerate}

To achieve a fully coupled global climate emulator, we must also predict sea ice. Inspired by the FMS coupler, we treat sea ice as the interface between the ocean's slow time step and the atmosphere's fast time step. SamudraI updates the sea ice concentration state every 5 days, in turn determining where ACE2 is allowed to prognose surface temperature over sea ice for the subsequent 5-day window. We use sea ice concentration -- defined as the sea ice grid cell fraction divided by (1 - land grid cell fraction) -- in order to avoid over-representing grid points with minimal ocean cover, and find that doing so reduces biases near coastlines where ocean fraction is small.
SamudraI was trained using the same constant learning rate schedule as ACE2 (Text S4). 
We show SamudraI pretraining loss and inference RMSE in Figure S1a.

\subsection{Fine-tuning SamudrACE}

Once ACE2 and SamudraI pretraining is complete, we select for coupled fine-tuning the checkpoint with the lowest normalized channel-mean root mean square error (RMSE) over all epochs for each respective model.
Together, the coupled emulators have a combined total of nearly 600 million parameters.
Prior to fine-tuning, initial coupled surface temperature biases are larger than the biases observed in either uncoupled model, especially in the ENSO region and above sea ice, as shown in Figure S16a.
Text S1 discusses this bias and its sensitivity to specific data processing choices which impact the consistency between uncoupled pretraining and coupled fine-tuning.

Coupled fine-tuning proceeds in two stages.
In the first stage, we fine-tune SamudraI by optimizing the loss of four forward steps (20 days) while coupled to the pretrained ACE2 model.
Throughout this first phase of coupled fine-tuning, ACE2's model weights are held fixed and atmospheric fields do not contribute to the training or validation loss.
However, as seen in Figure S1c, atmosphere fields do contribute to the ``best" checkpoint selection, which is the SamudraI checkpoint leading to the lowest normalized channel-mean RMSE across all ocean and atmosphere fields.
Figure S16b shows that the coupled surface temperature biases are greatly reduced after fine-tuning Samudra in coupled mode, with a reduction in global time mean RMSE from 2.265 to 0.355 K. 
Much of this improvement is realized after only a single epoch of fine-tuning.

In the second stage, we continue fine-tuning with the best SamudraI checkpoint found in stage one. 
Since ACE2 was not updated during the first phase, it is initialized from its best pretraining checkpoint. 
In contrast to the first stage, weights for both ACE2 and SamudraI are jointly optimized with a cosine-annealing decay schedule~\cite{loshchilov2016sgdr}. 
The simulation trajectory again spans four 5-day ocean steps, during which the atmosphere advances through a sequence of eighty 6-hour steps.
As before, we optimize the loss of all four ocean steps, but now we add the loss of the first two of the eighty atmospheric steps (Figure S1d).
We provide additional details of coupled fine-tuning hyperparameters in Text S4.

\subsection{Datasets}

Our reference training and evaluation datasets are from a 200-year preindustrial control simulation from GFDL's Climate Model v4 (CM4)~\cite{held2019structure},  rerun to save high frequency data, starting from year 151 of the original CMIP6 simulation. 
CM4 was run with a $\sim$100 km resolution C96 atmosphere with 33 terrain-following vertical levels and a 0.25 degree ocean on a tripolar grid with 75 hybrid pressure/isopycnal vertical coordinate levels. Both datasets were conservatively remapped to a 1 degree Gaussian grid with 8 terrain-following atmospheric layers and 19 ocean layers with constant-depth interfaces for emulator training and testing.
We use the first 155 years of output for training, the next 5 years for validation, and the remaining 40 years of data are held out for testing.
This training duration is supported by sensitivity analysis (Text S3; Figures S17 and S18) showing that longer records are essential for capturing internal variability like ENSO, even as time-mean error metrics stabilize with shorter datasets.

The reference AM4 atmospheric fields and SIS2 sea ice concentration were output fully consistently with ACE, with instantaneous snapshots of prognostic variables output every six hours. All surface and top-of-atmosphere fluxes (including precipitation) were output as six-hour time-averages between these snapshots.
This enables the surface fluxes to be accumulated over 20 atmospheric steps into 5-day averages suitable for forcing the ocean emulator.

The reference OM4 ocean fields were all output as 5-day averages, rather than instantaneous snapshots.  This includes 
the sea-surface temperature, used to force the ACE atmosphere model. For the purposes of this paper, these 5-day averages are used as an estimate of the ocean state at the midpoint of the averaging interval.

Achieving consistent land, ice, and open ocean masking between the emulators is a crucial step, with important implications for coupling.
Just as a traditional coupled GCM must reconcile differences in horizontal resolution when coupling the ocean to the atmosphere, GCM emulators must do so either during data preparation or via an online procedure.
In this work, we emulate CM4 with a common 1 degree horizontal resolution for both the atmosphere and ocean and therefore handle mask alignment during the data preparation step.
Early versions of the coupled emulator suffered from relatively large surface temperature biases on grid cells with coastal sea ice due to small inconsistencies in the sea ice from the separate atmosphere and ocean data processing pipelines which resulted in the use of inconsistent sea ice during uncoupled pretraining of Samudra and ACE.
These errors were easily corrected by instead pretraining Samudra with the sea ice resulting from the atmosphere data processing pipeline. We note that coupled fine-tuning of ACE was also an effective way to reduce these biases, allowing ACE to adapt to Samudra's sea ice representation in coupled mode.
In Text S1 we provide additional discussion of our data processing choices and their effects on the coupled emulator.

\section{Results}

SamudrACE can stably emulate CM4 with low bias in the time series of global means and time mean spatial error patterns over a held-out 40-year inference period covering the final part of the simulation. 
These biases are of similar magnitude to the biases of the uncoupled component emulators.
Using a single NVIDIA H100 GPU, a SamudrACE simulation generates around 1500 SYPD (6.4 days per second), a 3750$\times$ decrease in energy usage when compared to the CM4 simulation at 16 SYPD (0.068 days per second) using 6399 CPU cores on AMD EPYC 7H12 processors.
Beyond the brief summary provided
in Sections \ref{sec:ts-precip} --\ref{sec:ice}, we provide a comprehensive accounting of bias and other details in the Supplementary Material, similar to previous analyses of ACE2 and Samudra. 

\subsection{Climate mean state}
\label{sec:ts-precip}

In general, SamudrACE is a highly skilled emulator of climate mean states, not unlike its component ML emulators on their own.
SamudrACE's overall time mean state is accurate in terms of the RMSE of the generated time mean (Figure S2) and globally averaged annual mean series (Figures S3 and S4).

\begin{figure}
\noindent\includegraphics[width=\textwidth]{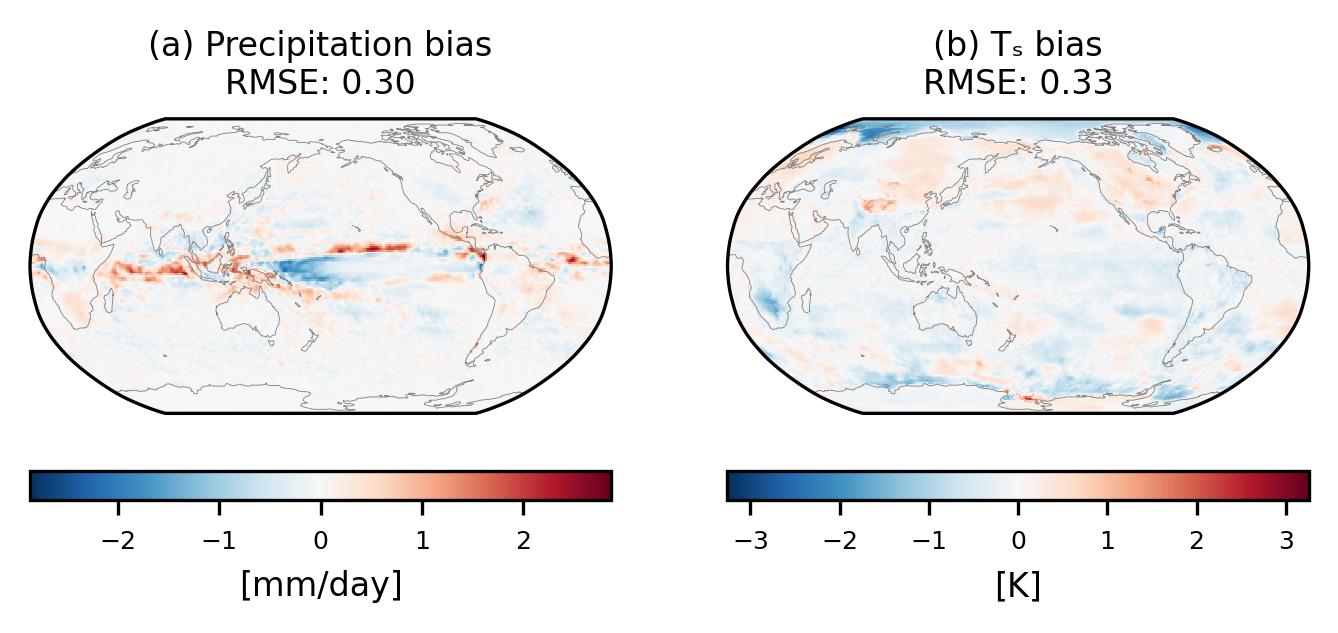}
\caption{Spatial bias patterns of the generated 40-year time-mean precipitation and surface temperature, computed as the difference between the time mean of the emulated output and the time mean of the reference simulation over the held-out inference period.}
\label{fig:sfc_temp}
\end{figure}

Figure \ref{fig:sfc_temp} shows that SamudrACE's atmosphere component faithfully reproduces CM4's time-mean precipitation and surface temperature over the 40-year held-out test period.
Time mean biases are small for both precipitation and surface temperature around the globe, with absolute maxima of just over 2 mm/day or 3 K, respectively, and small RMSEs.
Precipitation biases are concentrated in the tropics while surface temperature biases are pronounced in areas of sea ice and topographic features.
SamudrACE also accurately captures the statistical distribution of daily precipitation (Figure S15). 
We note the caveat that SamudrACE slightly underestimates extreme precipitation, a typical flaw in emulators that, like SamudrACE, are trained to minimize the mean squared error (MSE) loss \cite{perkins2025hiro}.

The ocean component similarly shows low time-mean spatial bias patterns, such as in the topmost layer currents and in the vertical structure of near-surface zonal sea water currents (Figures S5 and S6).
Figure S22 provides additional time-mean bias maps for sea surface height, column-integrated ocean heat content, and column-integrated ocean salt content.

The Atlantic Meridional Overturning Circulation (AMOC) is another important measure of time mean state on climate timescales. 
As a critical pathway for carrying warm water into the North Atlantic, the AMOC plays a vital role in regulating the global climate.
Figure S12(a) shows that SamudrACE emulates the vertical structure of CM4's AMOC with low bias in the 200-year time mean state.
Figure S12(b) gives the corresponding time series of AMOC anomalies in uncoupled SamudraI, SamudrACE and CM4.
Whereas uncoupled SamudraI lacks sufficient variability in the AMOC anomaly series, SamudrACE shows similar low frequency variability to CM4. 

While neither of SamudrACE's components guarantee exact closure of global energy budgets, they nonetheless generate budgets that are consistent with the reference simulation.
As seen in Figure S4, global ocean heat content in SamudrACE is stable over time and does not drift in a way that is inconsistent with CM4 and SamudrACE's meridional ocean heat transport shows encouraging agreement to the reference (Figure S20).
In the atmosphere, SamudrACE maintains a stable global energy budget with net energy fluxes and energy budget residuals that track closely with the reference CM4 simulation (Figure S14).
This combination of low bias and stability indicates that SamudrACE may be amenable to additional physical constraints such as energy budget corrections.

\subsection{Sea ice climatology}
\label{sec:ice}
 
\begin{figure}
\noindent\includegraphics[width=\textwidth]{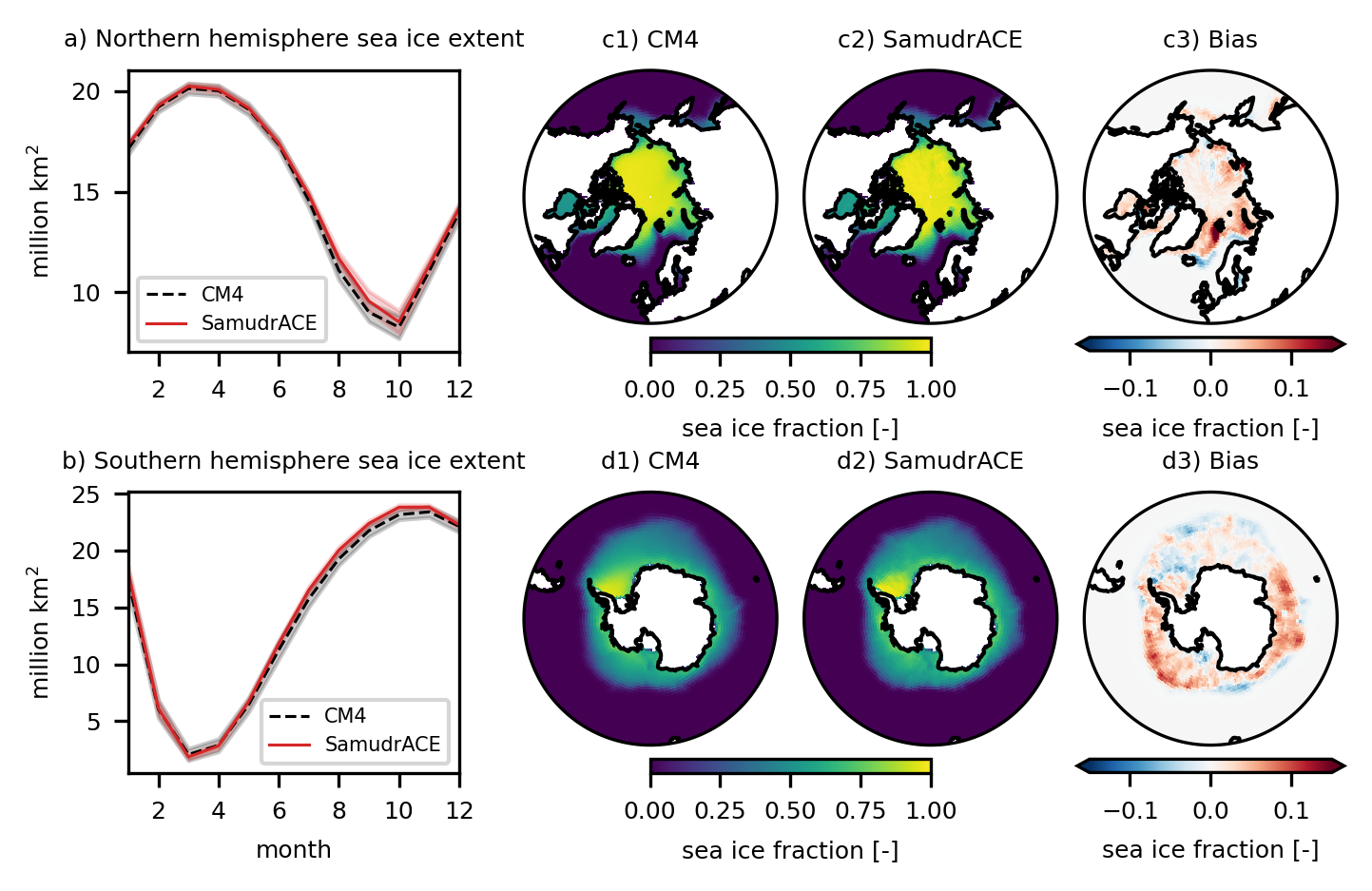}
\caption{Monthly mean over the 40-year held-out period of (a) Northern and  (b) Southern Hemisphere sea ice extent. Shading denotes the interannual standard deviation over 40 years. Panel c-d) shows the time mean sea ice fraction over the same time period for the CM4 target, SamudrACE, and its bias.
}
\label{fig:sea-ice}
\end{figure}

Overall, SamudrACE predicts a stable and accurate sea ice climatology with a realistic seasonal cycle. Sea ice fraction shown here is derived from SamudrACE's native sea ice concentration described in Section \ref{section:pretrain-ocean}. Figure \ref{fig:sea-ice}a-b shows monthly Northern and Southern hemispheres sea ice extent averaged over the 40-year held-out period. 
SamudrACE skillfully simulates the CM4 seasonal cycle with minimal bias.
Figure \ref{fig:sea-ice}c-d) shows the time-mean sea ice fraction for the Arctic and Antarctica for the CM4 target, SamudrACE, and the bias. 
SamudrACE tends to have a larger bias in sea-ice concentration near the edge of where sea ice exists. 
Notably, the highest positive bias is in the Greenland Sea, corresponding to where SamudrACE also simulates a weaker surface wind stress (not shown). 
We provide a similar view of sea ice thickness in Figure S13.

\subsection{ENSO}
\label{sec:enso}

\begin{figure}
\noindent\includegraphics[width=\textwidth]{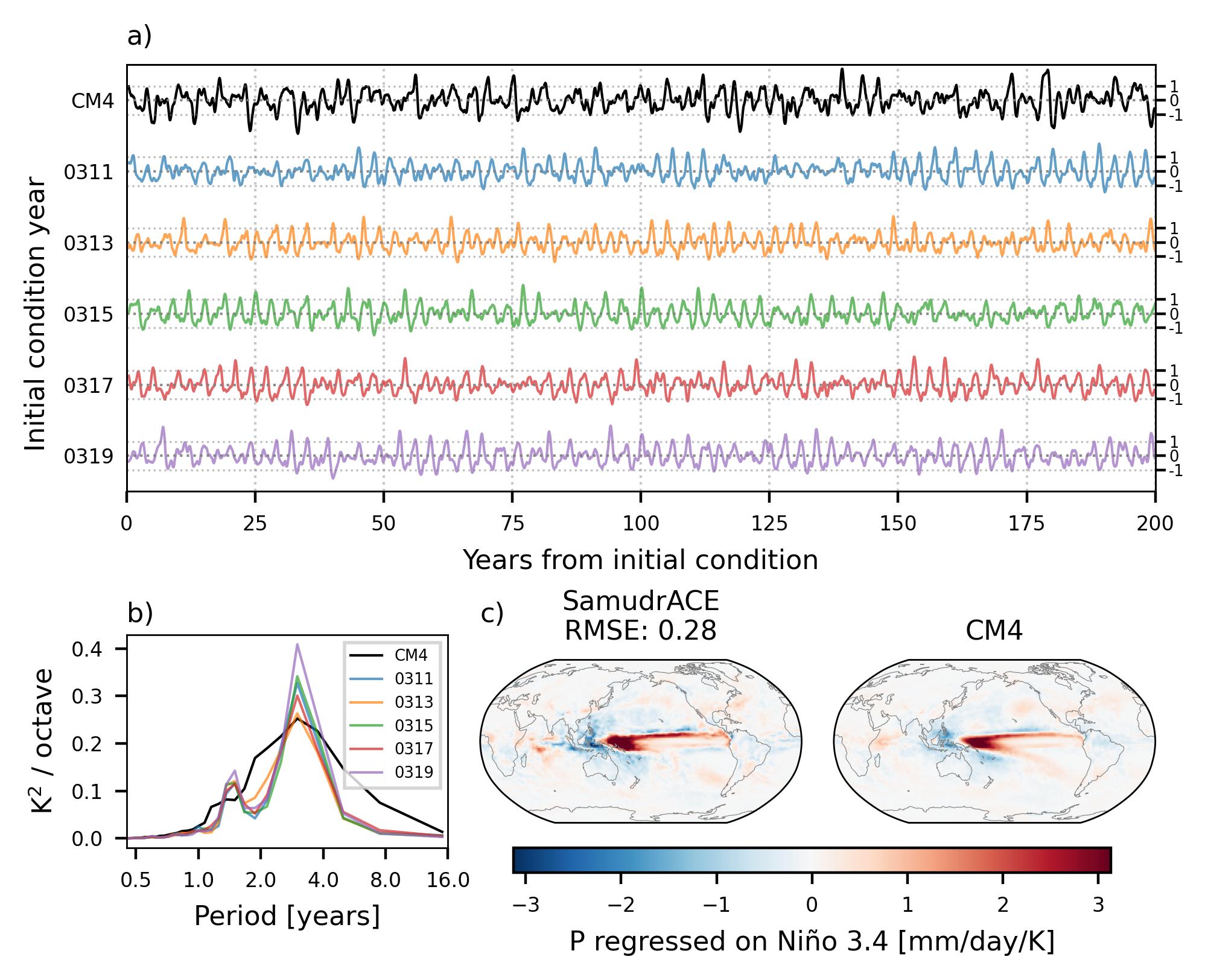}
\caption{ENSO characteristics in the 200-year CM4 simulation (black) and 5 separate rollouts of SamudrACE starting from different initial conditions (colors):
(a) time series of monthly mean Niño 3.4 index; 
(b) corresponding temporal power spectra (K$^2$/octave) computed using Welch's method with 15-year segments~\cite{welch1967use,harris2005use};
(c) regression of the spatial pattern of precipitation on the Niño 3.4 index in CM4 and SamudrACE for the held-out 40-year inference period.
}
\label{fig:enso}
\end{figure}

Figure \ref{fig:enso} compares ENSO characteristics in CM4 and SamudrACE. The emulator captures fundamental ENSO dynamics, including large-amplitude events, spatial variability patterns, and the characteristic 3-year spectral peak. Sensitivity analysis (Figure S18) confirms that the full training volume is necessary for Niño 3.4~\cite{barnston1997documentation} amplitude to converge toward the reference.
In 200-year rollouts initialized from five different years (Figure \ref{fig:enso}a), SamudrACE exhibits comparable but slightly weaker variability than CM4. Specifically, the emulator shows a bias toward sharper El Niños (max 2.1~K) and weaker La Niñas (min -1.6~K vs -2.4~K in CM4), and occasionally exhibits decade-long periods of suppressed activity. This asymmetry is further detailed in Figure S7.
Spectral analysis (Figure \ref{fig:enso}b; see Text S2 for methods) reveals that while SamudrACE reproduces the 3-year peak, it generally exhibits excessive power in the 2--4 year band and a deficit at lower frequencies ($>$ 4 years). This high-frequency bias is corroborated by unrealistically high autocorrelation at 36-month lags (Figure S10).
Teleconnections are also well-represented: the regression of precipitation onto the Niño 3.4 index matches the reference (Figure \ref{fig:enso}c). Comparison with uncoupled ACE2 baselines confirms that ACE2's intrinsic response to SSTs remains largely unchanged by coupled fine-tuning (Figure S19). Furthermore, despite weaker La Niñas, the vertical structure of equatorial Pacific temperature anomalies remains realistic across ENSO phases (Figures S8 and S9).

\subsection{Decadal coupled variability and the Interdecadal Pacific Oscillation}

The Interdecadal Pacific Oscillation (IPO) is a pattern of Pacific Ocean variability characterized by multiyear-to-decades-long periods of sustained anomalous SSTs, with important implications for wildlife ecology\cite{mantua1997pacific}. Like ENSO, the IPO is a key feature of emergent coupled variability, though at a much lower periodicity. SamudrACE's generated IPO Tripolar Index~\cite{henley2015tripole} has a clear pattern of sustained anomalous SST with transitions between extreme polarities on timescales that are similar to those seen in CM4 (Figure S11). However, consistent with the spectral analysis of ENSO, the magnitude of this low-frequency variability is generally weaker than in the CM4 reference.

\section{Conclusions}
SamudrACE demonstrates that coupling 3D atmosphere and ocean emulators can produce stable, centuries-long simulations with realistic emergent variability.
While promising, our analysis also reveals areas for future improvement. 
In ENSO and other coupled phenomena such as the IPO, the emulator generally underestimates low-frequency variability on time scales longer than 4 years.
The exact cause is unknown, but we hypothesize that one possible way to overcome this issue is by increasing the amount of training data, thereby allowing for greater sampling of low-frequency dynamics (Text S3; Figure S18).
However, that same analysis suggested that increasing the training data size leads to diminishing returns, so
future efforts could also focus on refining the fine-tuning strategy or model architecture to improve the coupled emulator's ability to learn low-frequency characteristics of the reference.

In our approach which couples the pretrained component emulators in physical state space, minor differences in the separate data processing pipelines for uncoupled pretraining of the atmosphere and ocean emulator become readily apparent in the coupled emulator. 
This is both an advantage and a pitfall of our uncoupled-pretraining-to-coupled-fine-tuning pipeline combined with physics space coupling.
Skipping the uncoupled pretraining stage may help to avoid this issue by directly optimizing randomly initialized component emulators in coupled mode, though we found this approach to be computationally inefficient in early experiments.
Alternative coupling approaches in component emulator latent space may also be able to circumvent these issues, at the cost of lowered physical interpretability and obfuscation of inconsistencies in uncoupled pretraining.
Finally, it may also be possible to avoid offline horizontal regridding of ocean and sea-ice outputs entirely by direct emulation of the native-resolution global ocean model outputs and online deterministic or learned regridding when coupling to an atmosphere emulator.

While SamudrACE reproduces the general distribution of daily precipitation (Figure S15), it slightly underestimates extremes due to the exclusive use of MSE loss, a known limitation of the ACE2 component \cite{perkins2025hiro}. Future versions aim to mitigate this over-smoothing by incorporating stochastic training with CRPS or spectral energy score losses, enabling longer autoregressive trajectories without sacrificing spectral fidelity.

The successful coupled GCM emulation framework of SamudrACE provides a clear pathway toward emulating a complete Earth system by incorporating additional components, such as land and biogeochemical models, opening new avenues for efficient, large-ensemble climate studies. 
As a natural next step, future work could explore the incorporation of a sea-ice emulator able to prognose sea ice concentration at the original 6-hourly temporal resolution used in forcing uncoupled ACE2, and for handling atmosphere-ocean flux and momentum exchange.

In its current configuration and training on a preindustrial control simulation, we do not expect SamudrACE to be capable of generalizing to unseen forced conditions, including abrupt jumps or steady increases in global CO2 concentrations.
However, this study validates the stability and skill of the coupled architecture under control conditions, establishing a foundation for future applications to forced scenarios.

\section*{Open Research Section}

Prior to publication, the source code for uncoupled pretraining, coupled fine-tuning, and evaluation of SamudrACE will be released in a future version of the open source repository \url{https://github.com/ai2cm/ace}.
The SamudrACE model weights artifact, initial conditions, and forcing data (incoming solar radiation and time-invariant inputs) are available at \url{https://huggingface.co/allenai/SamudrACE-CM4-piControl}.
The full processed 200-year CM4 simulation will be made publicly available.

\acknowledgments

Ai2 is supported by the estate of Paul G. Allen.  This research received support through Schmidt Sciences, LLC, under the M$^2$LInES project. 
We thank all members of the M$^2$LInES team for helpful discussions and their support throughout this project.
This research was also supported in part through the NYU IT High Performance Computing resources, services, and staff expertise.  We also thank members of the NOAA Geophysical Fluid Dynamics Laboratory and Princeton University community for their support, specifically V. Ramaswamy for allocating GFDL computing resources for the CM4 reference simulation, Mitch Bushuk, Huan Guo, Pu Lin, and Sergey Malyshev for their help as we discussed which diagnostics to output from the CM4 run, and Linjiong Zhou for constructive comments on an earlier draft of this manuscript.
Lastly, we thank Nathaniel Cresswell-Clay and co-authors of \citeA{cresswell2024deep} for sharing SST outputs from the DLESyM coupled atmosphere and SST emulator.
We acknowledge the use of AI tools in the preparation of this work and manuscript, specifically, for code generation and for stylistic edits of the manuscript draft for improved clarity and succinctness, using the following models: Google Gemini 2.5 Flash and 2.5 Pro, and Anthropic Claude Sonnet 4 and Opus 4.1.

\section*{Conflict of Interest}

The authors declare there are no conflicts of interest for this manuscript.

\bibliography{references}

\appendix

\setcounter{figure}{0}
\renewcommand{\thefigure}{S\arabic{figure}}

\setcounter{table}{0}
\renewcommand{\thetable}{S\arabic{table}}

\section*{Supporting Information for ``SamudrACE: Fast and Accurate Coupled Climate Modeling with 3D Ocean and Atmosphere Emulators"}

In this Supporting Information we include additional details of our CM4 reference data processing pipeline (Text S1), the Niño 3.4 power spectrum computation used in the main text Figure 4a (Text S2), an ablation of training dataset size for uncoupled ocean component pretraining (Text S3), and training hyperparameters (Text S4).
Tables \ref{table:variables_samudra} and \ref{table:variables_ace} provide descriptions of SamudrACE's physical variables in the ocean and atmosphere components, respectively.
Finally, we provide Figures S1 to S22, referenced in the main text.

\noindent\textbf{Text S1. Data Processing and Sensitivity Analysis}

The results presented in the main text utilize a data processing pipeline we will refer to as the ``misaligned" setup. In this configuration, the 6-hourly atmospheric snapshots and 5-day averaged ocean states were processed such that the sea surface forcing of the atmosphere was applied inconsistently during uncoupled versus coupled training. Specifically, the time-averaged windows of surface flux forcings were aligned to the endpoints of the ocean’s 5-day averaging window. This alignment strategy introduced specific unseen biases into the pretraining phase that manifested in coupled mode as a pattern of dipole errors in the surface temperature time mean bias maps with ENSO-like features, as seen in Figure \ref{fig:coupled-tm-bias-maps}a.

While the manuscript results are derived from the "misaligned" setup, the schematic in Figure 1 depicts a refined methodology which we will call the ``aligned" setup. This improved approach removes inconsistency between uncoupled pretraining and coupled fine-tuning by aligning the ocean SST and sea ice state to the midpoint of the original averaging interval. 
As depicted in Figure 1, this results in an offset of 2.5 days between the surface flux 5-day averaging windows and the 5-day averaging windows utilized in the CM4 reference simulation run when saving OM4's outputs. 
The ``aligned" setup effectively removes the large sea surface temperature biases seen in Figure \ref{fig:coupled-tm-bias-maps}a and greatly reduces the biases above sea ice. 
Moreover, this new setup yields slightly lower channel-mean global time-mean RMSEs even without the second phase of coupled fine-tuning in which both components are jointly optimized.
However, we retained the "misaligned" setup for the reported results due to observed trade-offs: the new setup resulted in minor degradation in the time-mean bias patterns for surface temperature and precipitation and exhibited slightly weaker Niño 3.4 variability on the held-out 40 year inference set.
In spite of these downsides, we believe that the ``aligned" setup is the correct approach for future versions of the coupled emulator.
Finally, we note that the correct choice of time alignment would be unambiguous if we had saved out snapshots for the ocean data rather than time means.

\noindent\textbf{Text S2. Niño 3.4 Power Spectrum Computation}

The Niño 3.4 power spectra in Figure 4b were computed using Welch's method \cite{welch1967use} with a segment length of 15 years (180 months) and default 50\% overlap between segments. The power spectral density (PSD) was converted to units of K$^2$/octave by multiplying by the frequency and $\ln(2)$, which weights the spectrum by the width of equal-ratio frequency bands and facilitates visual interpretation of power distributed across different timescales \cite{harris2005use}. We chose this particular spectral estimation method and segment length empirically to provide a reasonable visual match to the Niño-3 SST spectra presented in Figure 31 of \citeA{held2019structure}, which displays CM4.0's ENSO characteristics using a Morlet wavelet analysis with wave number 6. While our Welch method differs from the wavelet approach used by \citeA{held2019structure}, both methods capture the broad interannual ENSO peak spanning 2–8 year periods. The 15-year segment length provides sufficient frequency resolution to resolve ENSO's dominant periods while offering adequate averaging over the 200-year simulations to reduce spectral variance. This choice enables direct qualitative comparison of SamudrACE's ENSO spectral characteristics with those of the parent CM4.0 model as documented in the literature.

\noindent\textbf{Text S3. Ablation of training dataset size}

Figures S17 and S18 evaluate the sensitivity of SamudraI’s pretraining skill to dataset size, focusing on global time-mean RMSE and ENSO variability over a 40-year held-out period. We focus on the uncoupled SamudraI (forced with AM4 surface fluxes) because the full SamudrACE pipeline is computationally prohibitive, and the ocean’s slower evolution likely dictates the data requirements.

For each dataset size, we trained an ensemble of three models with random initialization. Figure S17 shows a reduction in global RMSE with increasing data; however, most variables (excluding potential temperature ``thetao") plateau near 80 years. Conversely, Figure S18 shows that ENSO variability improves significantly between 80 and 160 years, even as SST RMSE stabilizes. This suggests that while diminishing returns exist, sufficiently long records are required to adequately capture low-frequency variability like ENSO.

\noindent\textbf{Text S4. Training hyperparameters}

All models were trained using the AdamW optimizer with a batch size of 16.
The atmosphere model was trained for 50 epochs (707,200 gradient steps) with a constant learning rate of $1 \times 10^{-4}$.
The ocean model was trained for 150 epochs (106,050 gradient steps) with the same constant learning rate of $1 \times 10^{-4}$.
In Stage 1 of training SamudrACE we fine-tuned the ocean component for 20 epochs (14,140 gradient steps) using a constant learning rate of $1 \times 10^{-4}$.
In Stage 2, both components were jointly optimized for an additional 20 epochs (14,140 gradient steps) with a cosine-annealing decay schedule~\cite{loshchilov2016sgdr} initialized at $1 \times 10^{-5}$.

\begin{figure}
\noindent\includegraphics[width=\textwidth]{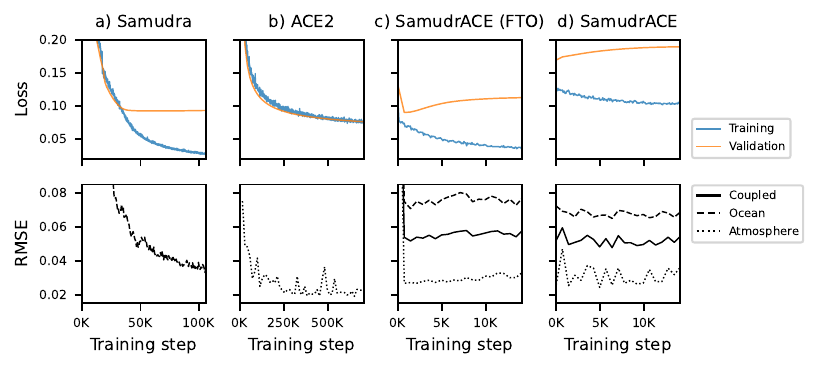}
\caption{
Training and validation loss and channel-mean RMSE.
Samudra and ACE2 are pretrained in uncoupled mode for 150 epochs (106,050 steps) and 50 epochs (707,200 steps), respectively, and both achieve minimum channel-mean RMSE and validation loss late in training.
After pretraining, ocean-only coupled fine-tuning (``SamudrACE (FTO)") was carried out for 20 epochs (14,140 steps), during which the Samudra checkpoint with the lowest RMSE was coupled to the pretrained and fixed ACE2 checkpoint with lowest RMSE.
At initialization the coupled pretrained emulators result in large validation loss and coupled RMSE (0.13 and 0.23, respectively, at training step 0). After 2 completed epochs the ocean-only coupled fine-tuning achieved its lowest coupled RMSE of 0.052, averaged over all ocean and atmosphere channels.
This checkpoint was then further fine-tuned for an additional 20 epochs, updating both Samudra and ACE2 (``SamudrACE"), and reached the lowest coupeld RMSE of 0.048 after 9 completed epochs.
RMSEs are averaged across all channels, where the ``Coupled" RMSE (solid black line) is the weighted average of the ``Ocean" (dashed black line) and ``Atmosphere" (dotted black line) channel-mean RMSEs in coupled fine-tuning runs.
}
\label{fig:training}
\end{figure}

\begin{figure}
\noindent\includegraphics[width=\textwidth]{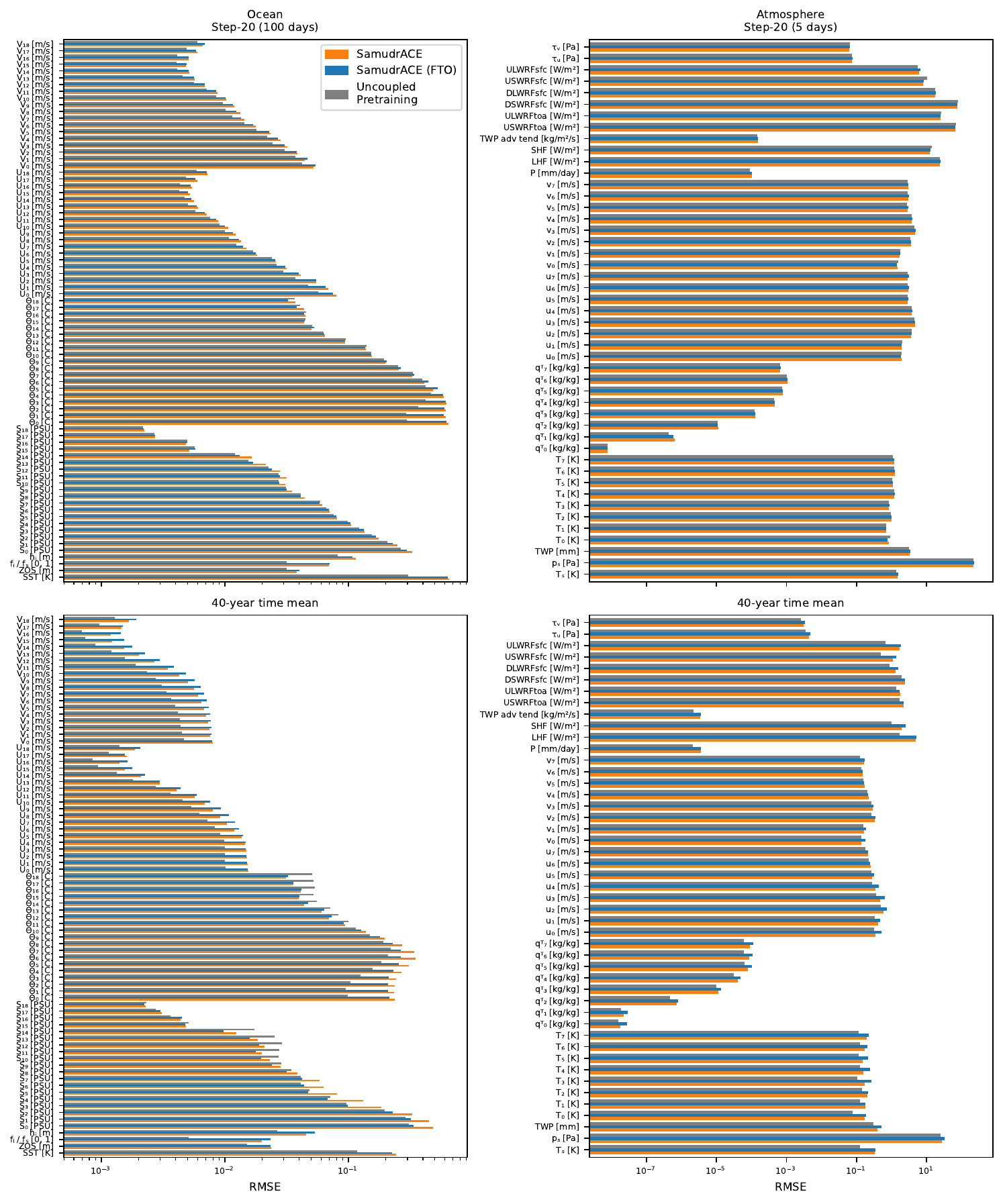}
\caption{Step-20 and 40-year RMSEs for the time-mean spatial structure of all oceanic and atmospheric output fields.}
\label{fig:time_mean_rmse}
\end{figure}

\begin{figure}
\noindent\includegraphics[width=\textwidth]{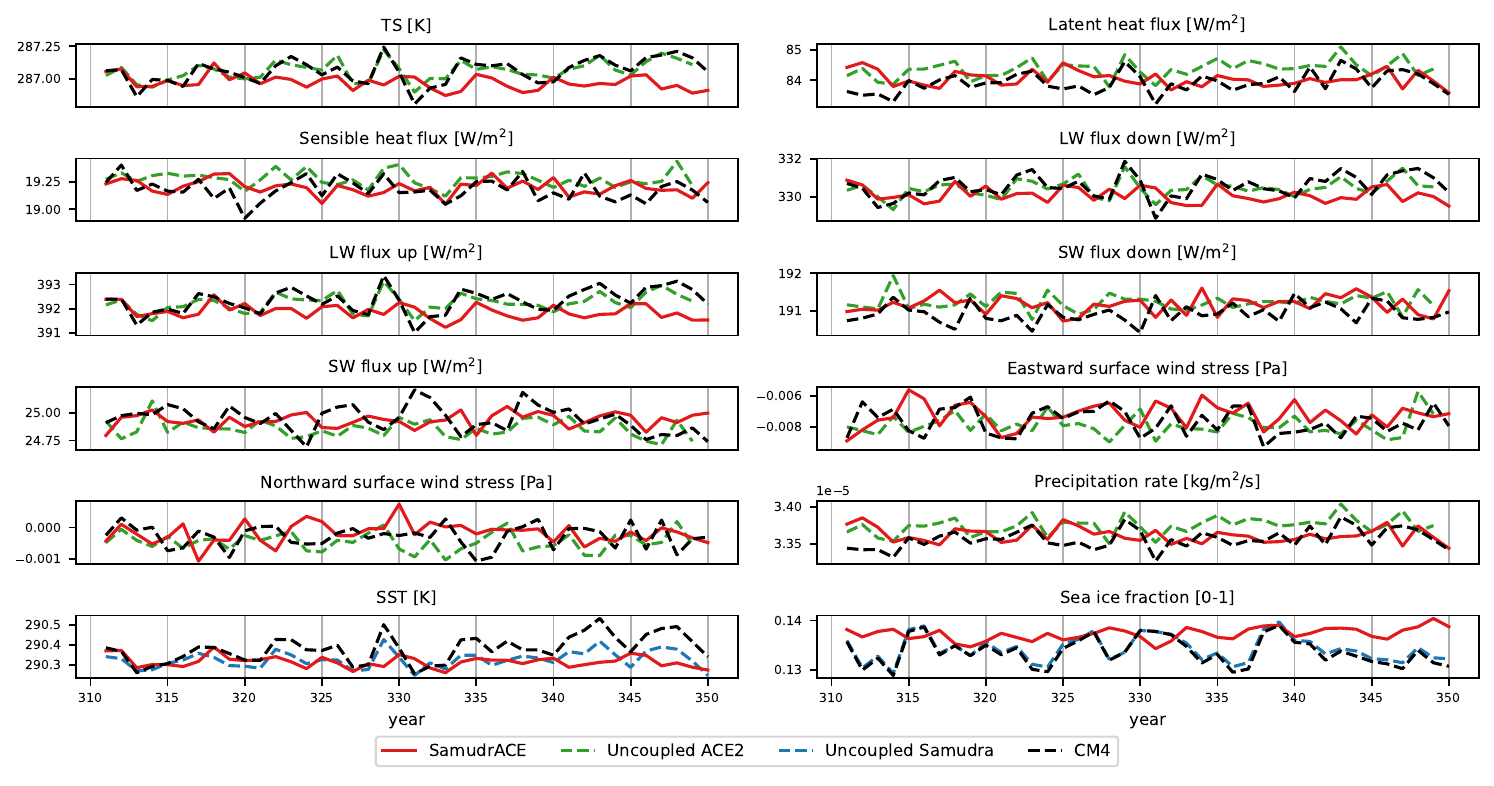}
\caption{
Annual and global mean time series over the 40-year test period for key surface variables involved in the exchange of information between the atmosphere and ocean components of the coupled model.
We show SamudrACE, uncoupled ACE2, uncoupled Samudra, and CM4 reference.
In uncoupled mode, the strong effect of prescribed boundary conditions is apparent for several variables, particularly $T_S$, longwave fluxes, $SST$, and sea ice fraction.
In coupled mode, the generated boundary conditions in SamudrACE quickly decorrelate from CM4 as the simulation progresses, allowing for increased internal variability and the emergence of coupled physical phenomena.
}
\label{fig:annual_mean_series}
\end{figure}

\begin{figure}
\centering
\noindent\includegraphics[width=0.7\textwidth]{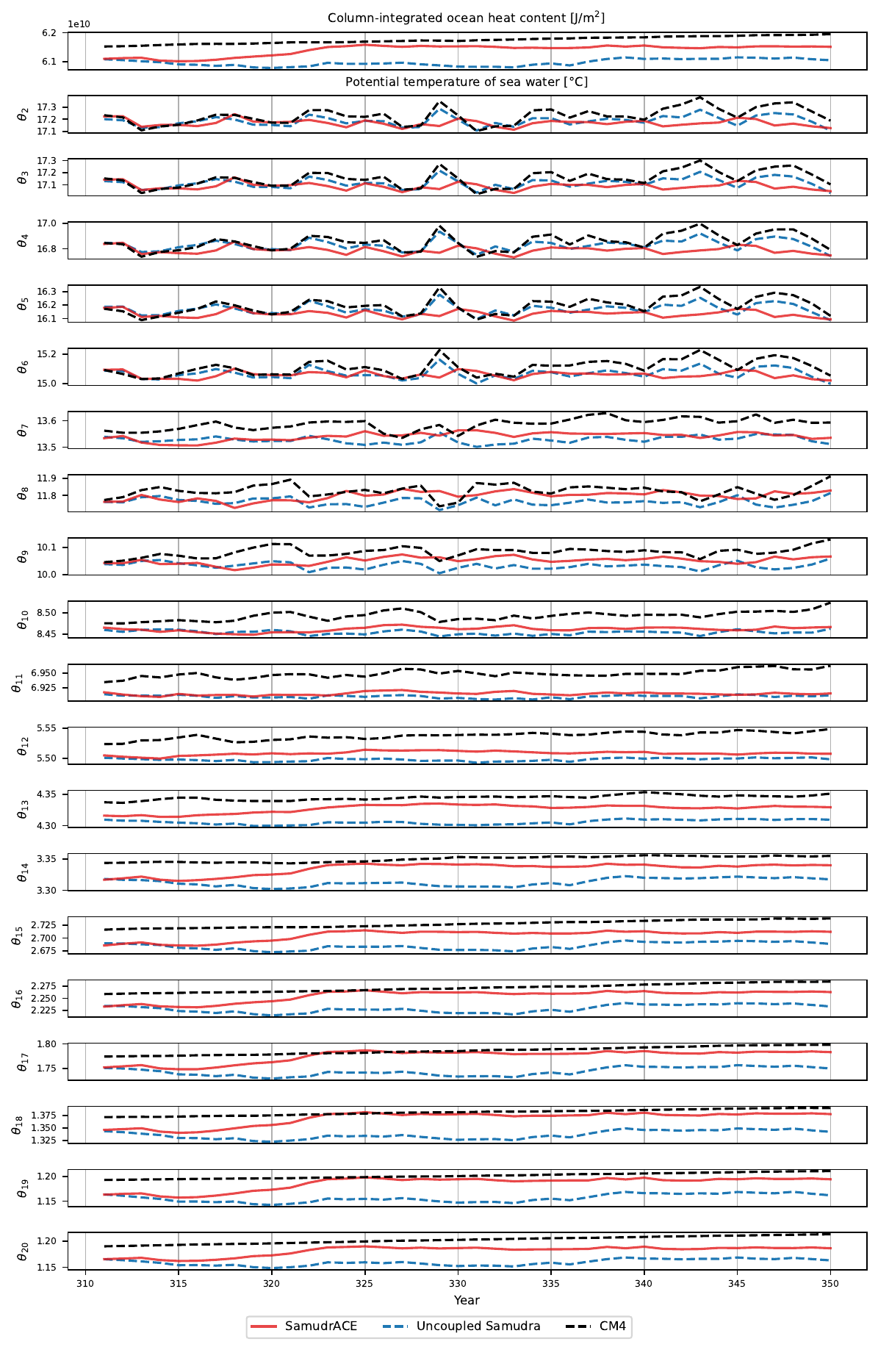}
\caption{Annual mean time series of column-integrated ocean heat content (OHC) and potential temperature at all 19 vertical ocean levels.
We show SamudrACE, uncoupled Samudra, and the CM4 reference.
SamudrACE substantially reduces the time-mean bias of OHC compared to uncoupled Samudra.
This improvement is primarily concentrated in the deeper ocean layers below a depth of 650 meters ($\theta_{o,10}$).
}
\label{fig:annual_mean_series_ohc}
\end{figure}

\begin{figure}
\noindent\includegraphics[width=\textwidth]{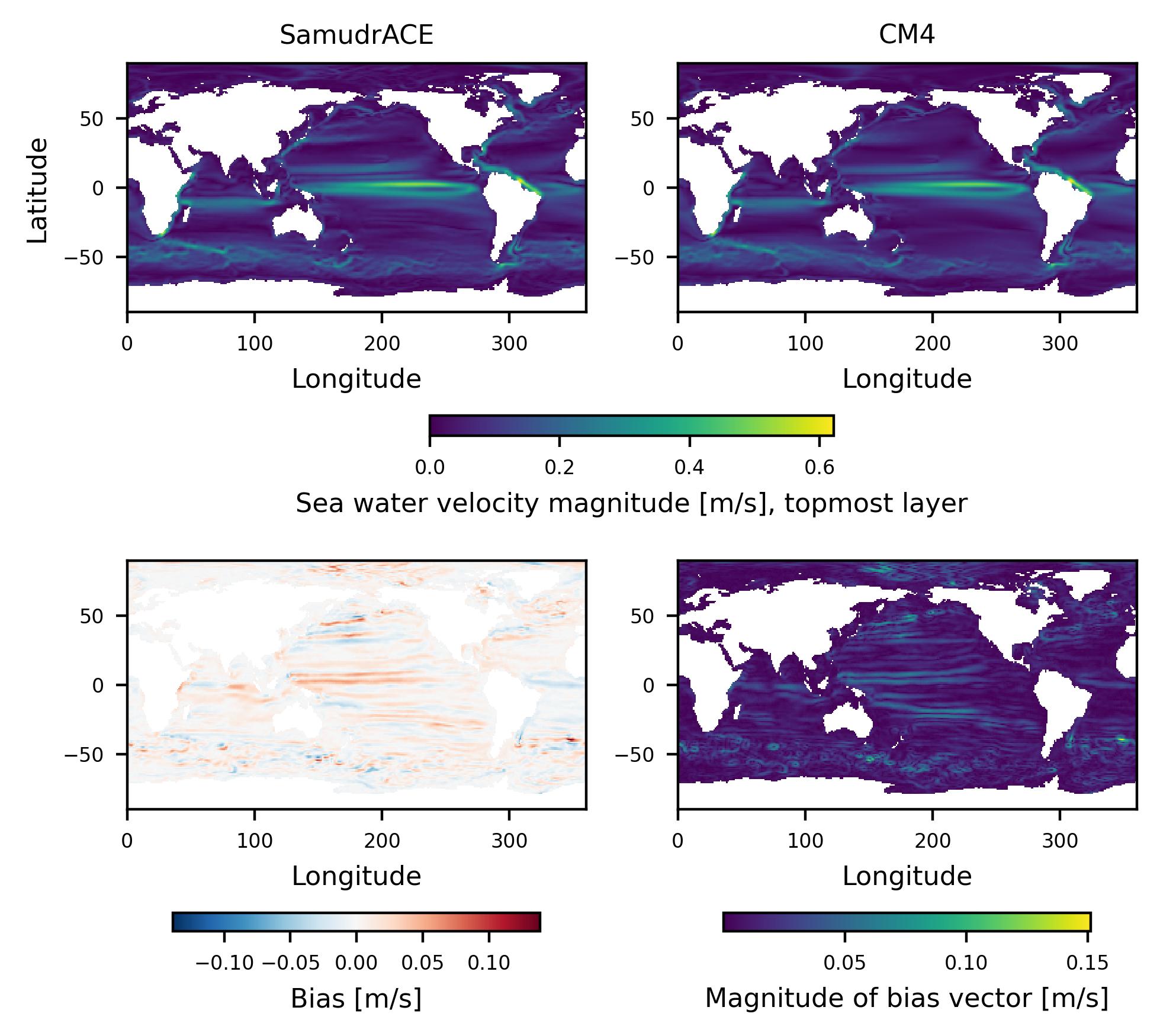}
\caption{
40-year time-mean generated and target sea water velocity magnitude for the coupled emulator and CM4 target (first row), together with the magnitude bias map (generated - target), and map of bias vector magnitude (second row).
Eddy-like bias patterns can be observed in the Southern Ocean, Northern Pacific, and North Atlantic.
}
\label{fig:sea_velocity_mag}
\end{figure}

\begin{figure}
\noindent\includegraphics[width=\textwidth]{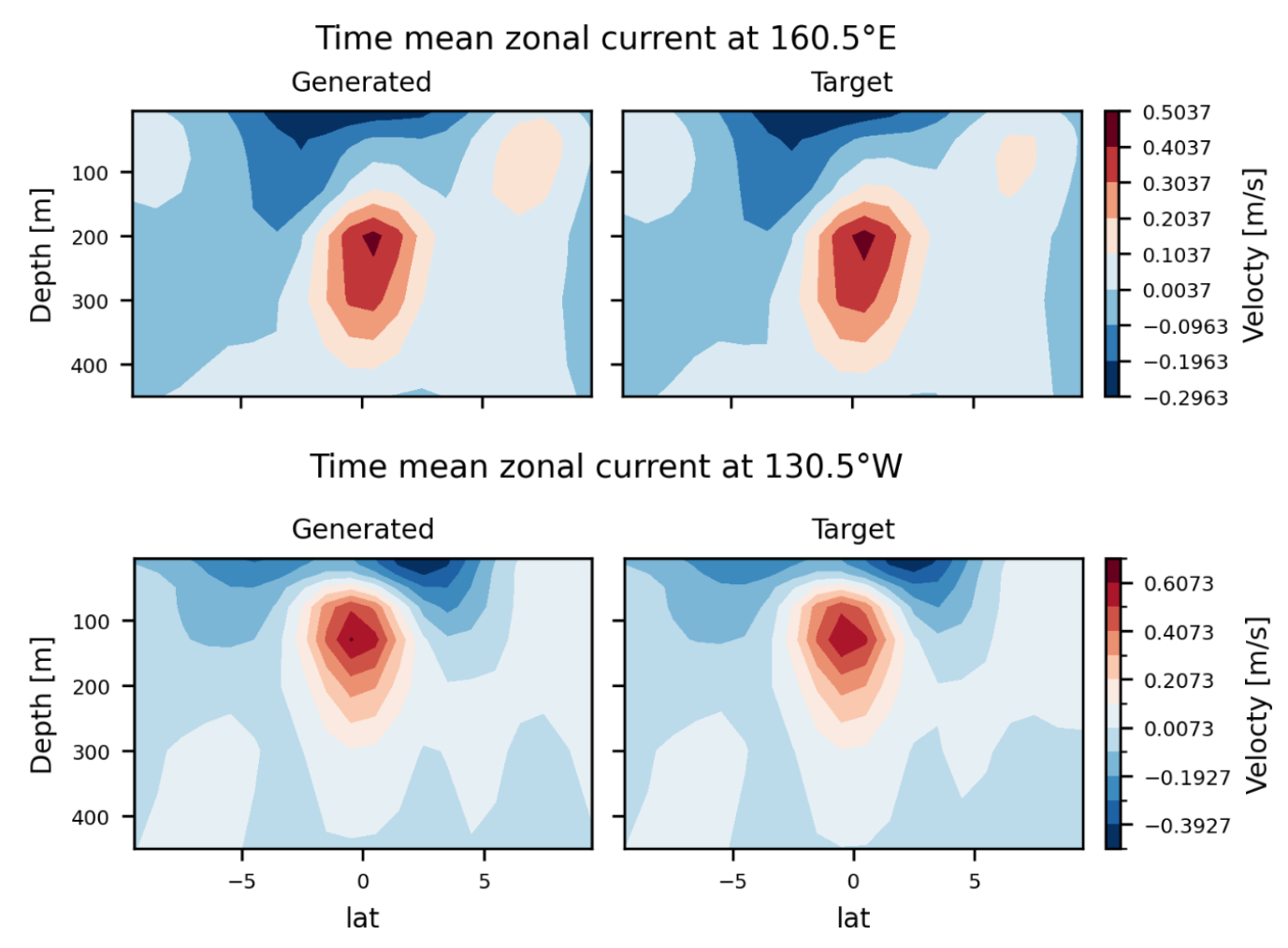}
\caption{Vertical and meridional structure of tropical zonal currents in the Pacific Ocean. 
Time-mean zonal sea water velocity from the surface to 400 meters depth for two vertical slices centered at the equator in the middle western Pacific at 160.5°E and 130.5°E. 
SamudrACE closely emulates the time average near-surface current in the equatorial Pacific.
}
\label{fig:currents}
\end{figure}

\begin{figure}
\noindent\includegraphics[width=\textwidth]{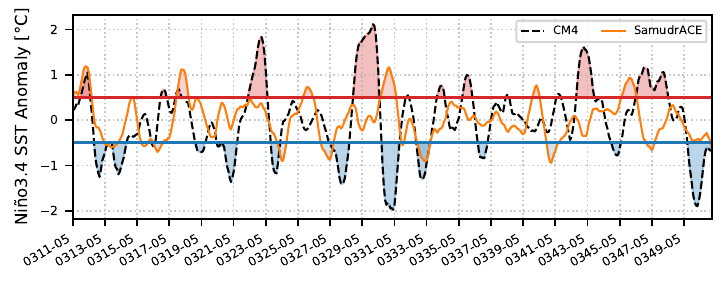}
\caption{Niño 3.4 SST anomalies over the 40-year held-out inference period.
The black dashed line is CM4 and the orange line is SamudrACE.
Events with anomalies of $\pm 0.5$ lasting longer than 5 months are highlighted with shading.
There were 6 El Niño and 7 La Niña events in the generated data and 9 El Niño and 11 La Niña events in the target data.
}
\label{fig:nino34_40yr}
\end{figure}

\begin{figure}
\noindent\includegraphics[width=\textwidth]{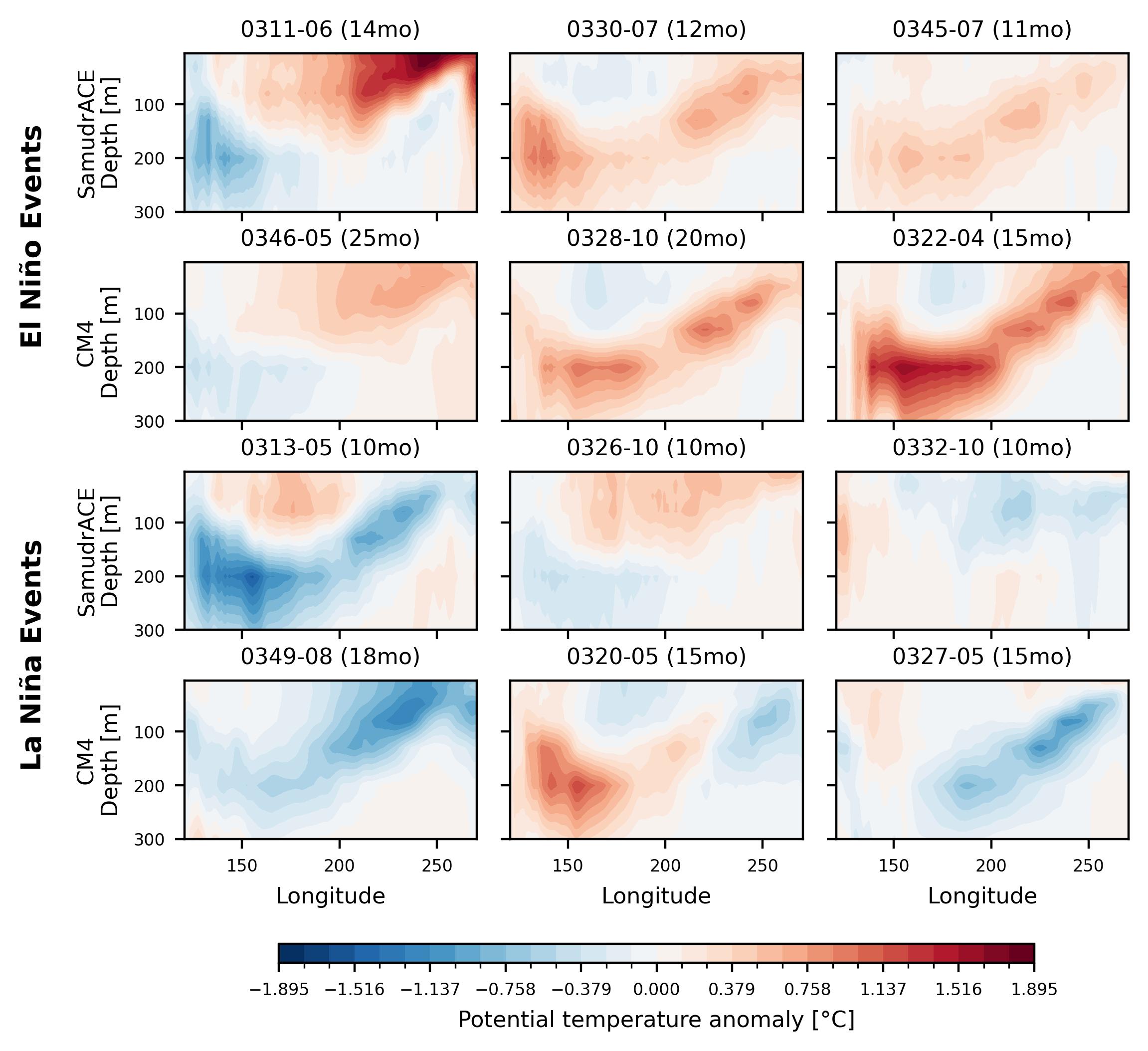}
\caption{
Time-mean potential temperature of sea water $\theta_o$ anomaly profiles in the equatorial Pacific.
For each model (SamudrACE and CM4) and ENSO condition (El Niño and La Niña) we select the top three events of Figure \ref{fig:nino34_40yr} in terms of duration, compute the $\theta_o$ anomaly for each month with respect to the 40-year time-mean, and visualize the time average taken over the duration of each event.
We give the event initialization and duration in months above each panel. 
}
\label{fig:thetao_enso_anomalies}
\end{figure}

\begin{figure}
\noindent\includegraphics[width=\textwidth]{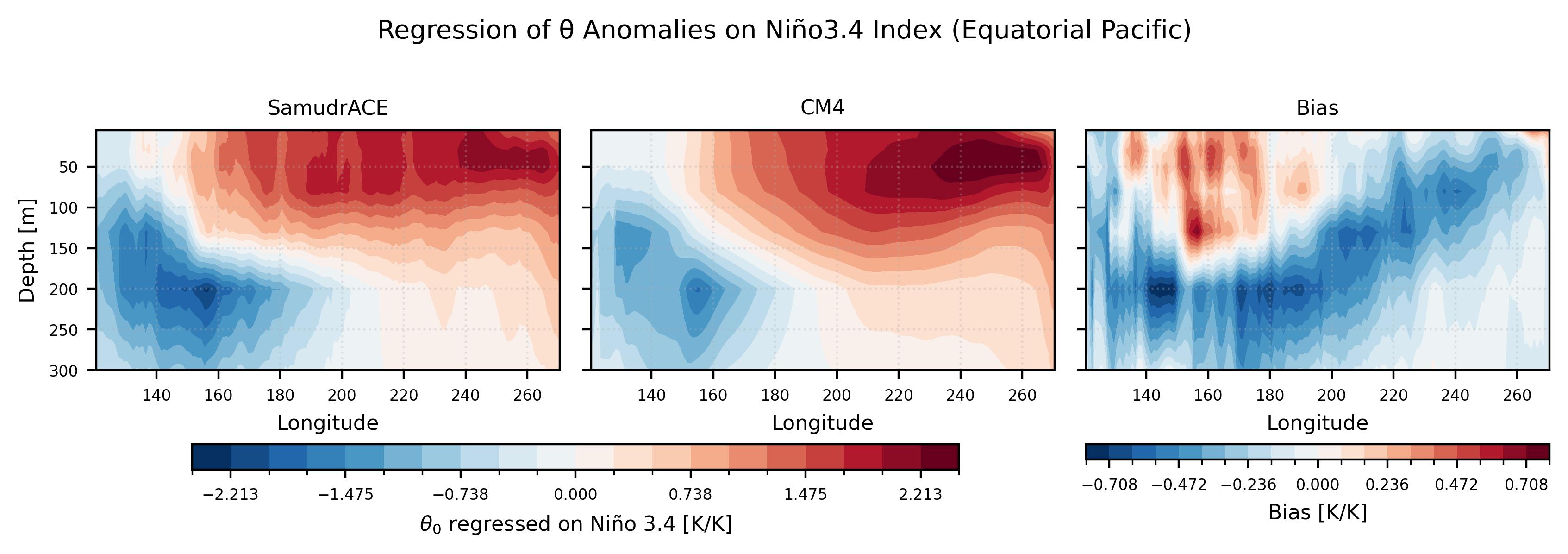}
\caption{
Regression of monthly mean $\theta_o$ anomalies on the Niño 3.4 index, respectively for generated and target outputs, over the 40-year held-out test dataset.
Biases in the response have a similar spatial structure to CM4's La Niña.
}
\label{fig:thetao_anomaly_regression}
\end{figure}

\begin{figure}
\noindent\includegraphics[width=\textwidth]{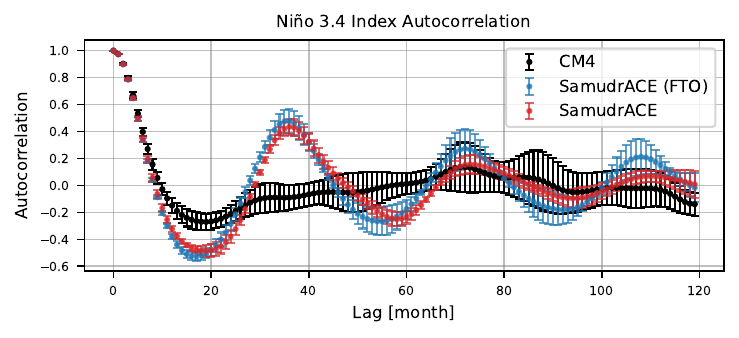}
\caption{Autocorrelation (lagged correlation) of the Niño 3.4 index for CM4 (black), SamudrACE (red), and FTO (blue). Uncertainty is expressed as the height of the bars above and below each dot, where the uncertainty is the standard deviation over the 5 ensemble members of SamudrACE and FTO. For CM4, we use a boot-strapping method where we randomly sample 25 different 500 month long segments from the single CM4 simulation. 
}
\label{fig:enso_autocorr}
\end{figure}

\begin{figure}
\noindent\includegraphics[width=\textwidth]{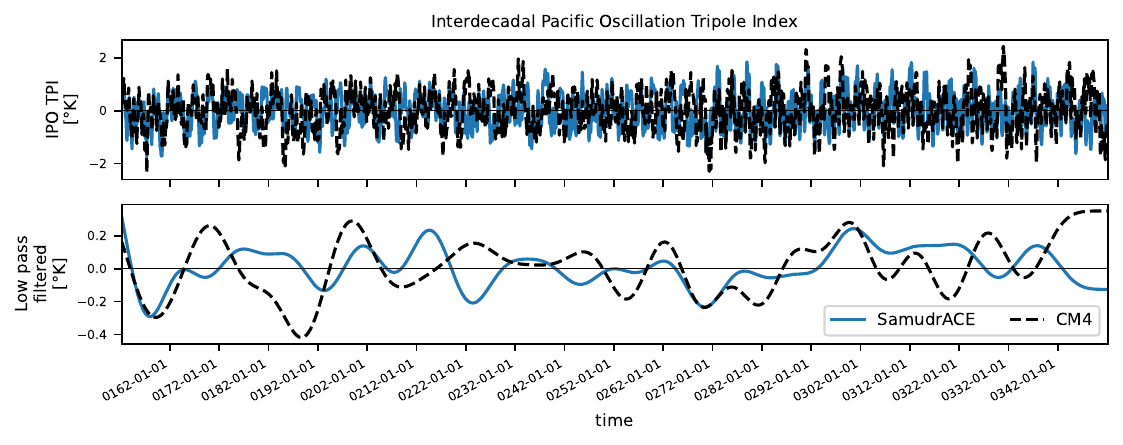}
\caption{
Interdecadal Pacific Oscillation Tripolar Index~\cite{henley2015tripole}.
The upper panel shows the unfiltered IPO TPI. 
Following \citeA{henley2015tripole} we apply a low-pass filter with cutoff of 13 years to extract decadal variability, shown in the lower panel.
The filter is Chebyshev Type I of order 5 and passband ripple of 0.5 dB.
We apply the filter in both the forward and reverse directions, resulting in a filtering with zero phase response and squared amplitude response.
}
\label{fig:tpi}
\end{figure}

\begin{figure}
\noindent\includegraphics[width=\textwidth]{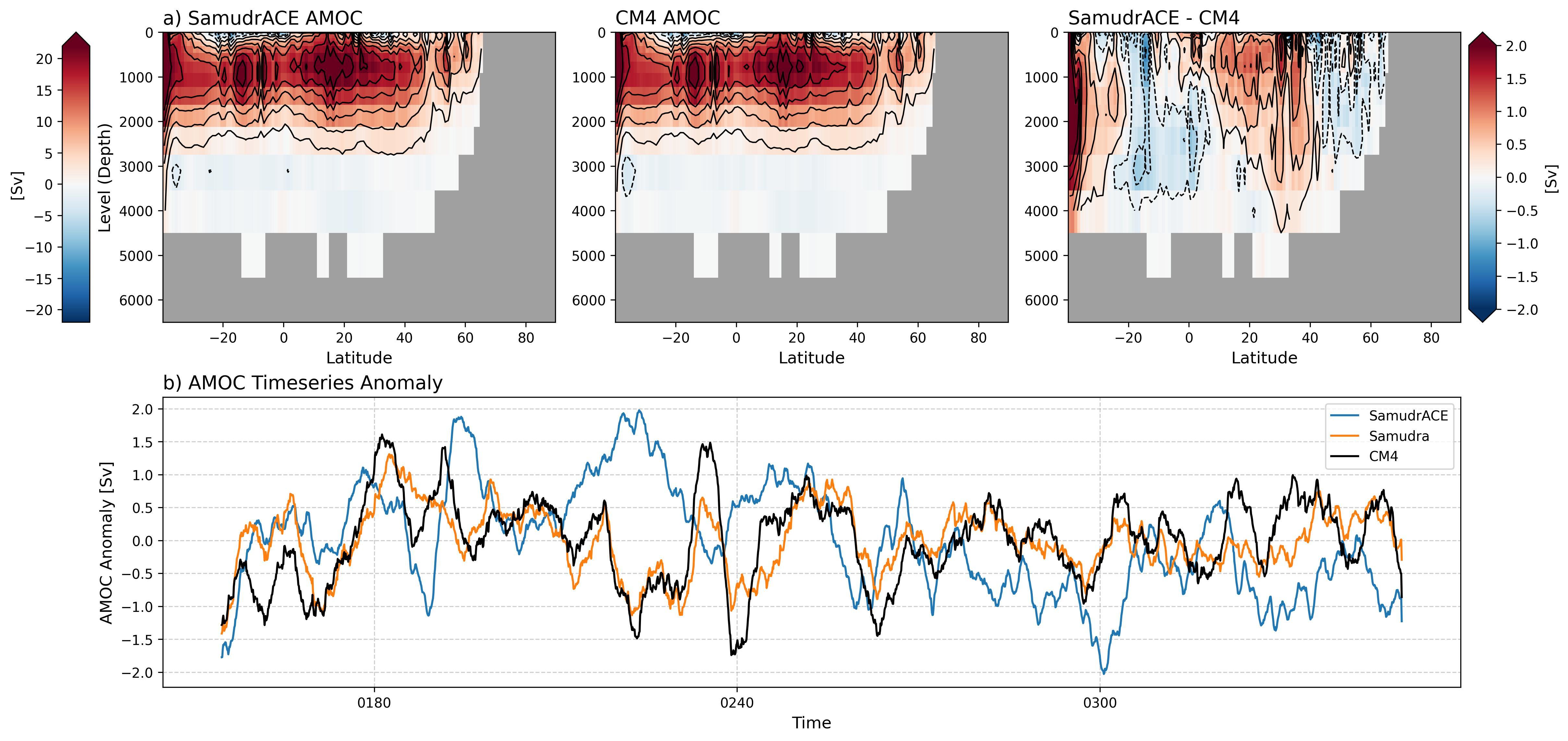}
\caption{Atlantic Meridional Overturning Circulation from a 200-year rollout in SamudrACE and CM4. Panel a shows the time mean of the AMOC generated with SamudrACE (left), the reference CM4 (middle), and the bias between the two (right). Panel b shows the time series of the AMOC strength anomaly, computed as the deviation from seasonal climatology of the maximum of the AMOC streamfunction between $28-32^\circ N$. Before computing the anomaly, we first apply a 5-year rolling mean to filter out high frequencies from the timeseries. For this panel we show the timeseries from CM4, SamudrACE, and for uncoupled Samudra prior to coupled finetuning and using CM4 boundary conditions (final 40 years). When computing the AMOC, we take the depth integral from the bottom of the ocean up to the top and include the contribution from dynamic sea level at the upper ocean cell. }
\label{fig:amoc}
\end{figure}

\begin{figure}
\noindent\includegraphics[width=\textwidth]{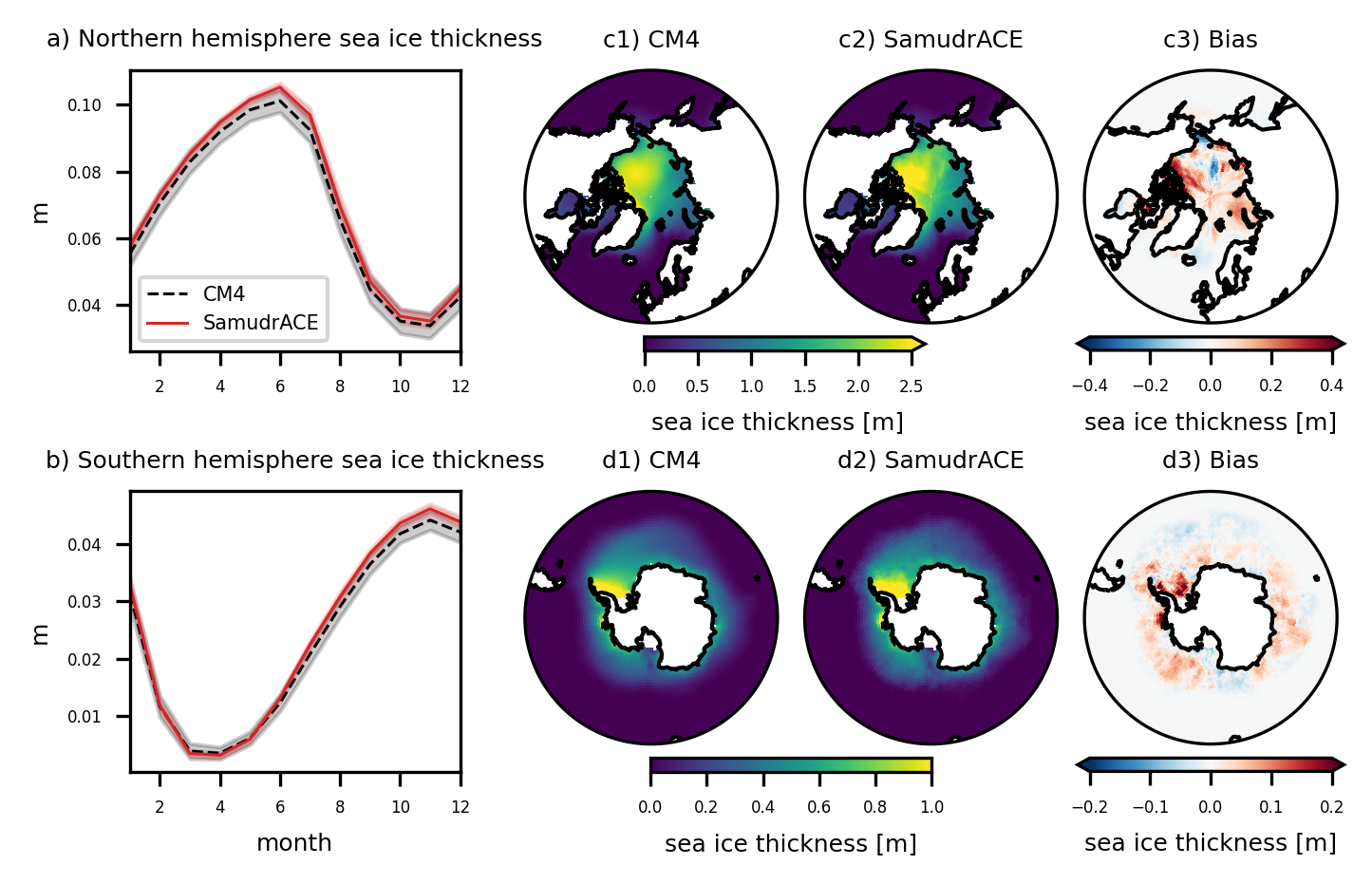}
\caption{Monthly mean over the 40-year held-out period of (a) Northern and  (b) Southern Hemisphere sea ice thickness. Shading denotes the interannual standard deviation over 40 years. Panel c-d) shows the time mean sea ice thickness over the same time period for the CM4 target, SamudrACE, and its bias.}
\label{fig:sea-ice-thickness}
\end{figure}

\begin{figure}
\centering\includegraphics[width=\textwidth]{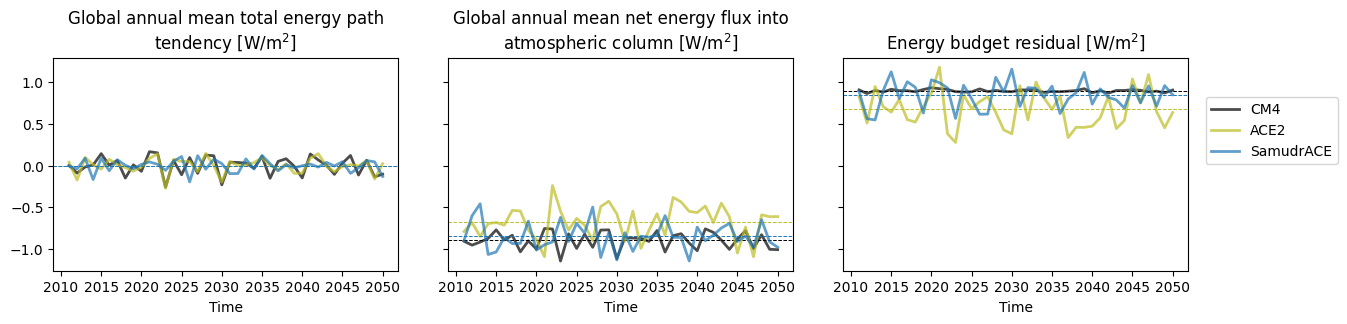}
\caption{Global energy budgets of the atmosphere component of CM4, ACE2, and SamudrACE. The energy budget residual is equal to the total energy path tendency minus the net energy flux into the atmospheric column. Note that the CM4 reference has time mean non-closure of around 0.9 W/m$^2$ due to the omission of total frozen precipitation rate as part of the net energy flux into the atmosphere column, since we did not include it as a diagnostic in this version of ACE2 and SamudrACE.}
\label{fig:implied_advective_tendency}
\end{figure}

\begin{figure}
\centering\includegraphics[width=0.8\textwidth]{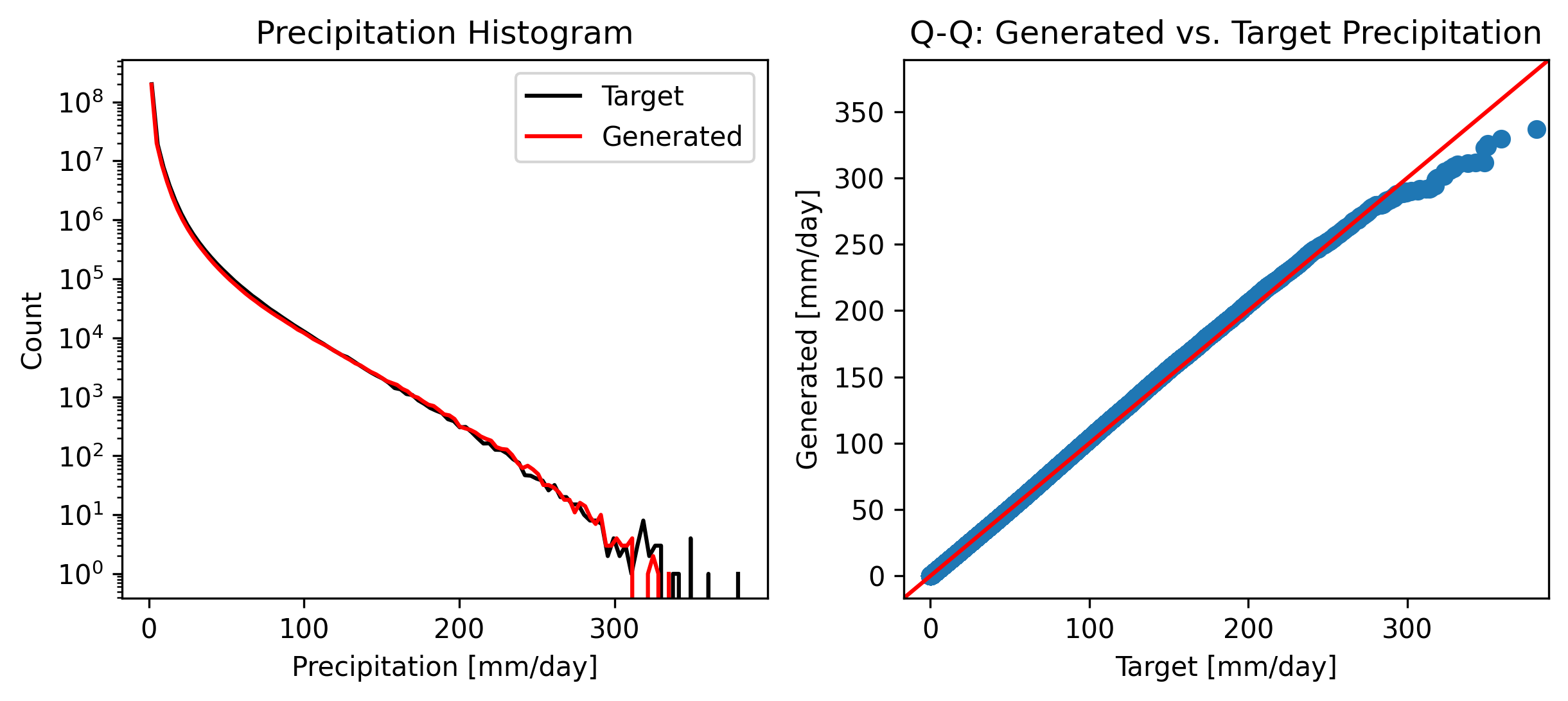}
\caption{Daily mean precipitation histogram (left) and quantile-quantile (right) comparison over the first 10 year of the test period.}
\label{fig:precip-hist-qq}
\end{figure}

\begin{figure}
\centering\includegraphics[width=0.8\textwidth]{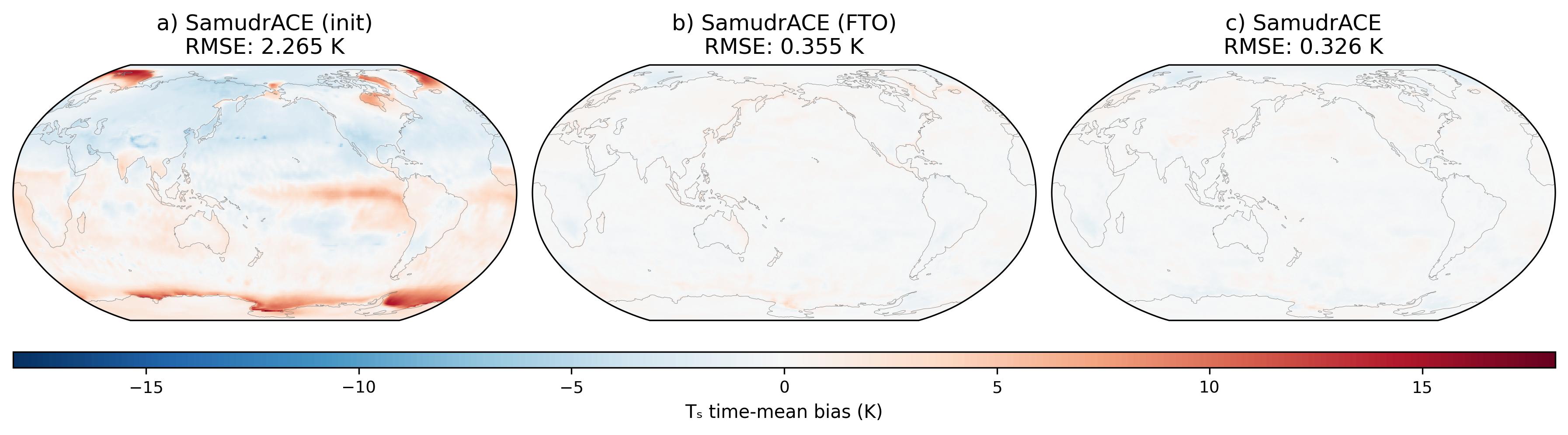}
\caption{40-year time-mean bias map of surface temperature at three points in the coupled fine-tuning trajectory: a) after initializing SamudrACE with the uncoupled pretraining weights from ACE and Samudra, prior to any coupled fine-tuning; b) after fine-tuning the Samudra component of SamudrACE to the point at which SamudrACE reached its lowest coupled inference error in the first phase of coupled fine-tuning; and c) after joint coupled fine-tuning of both components to the point of lowest overall coupled inference error.}
\label{fig:coupled-tm-bias-maps}
\end{figure}

\begin{figure}
\centering\includegraphics[width=0.8\textwidth]{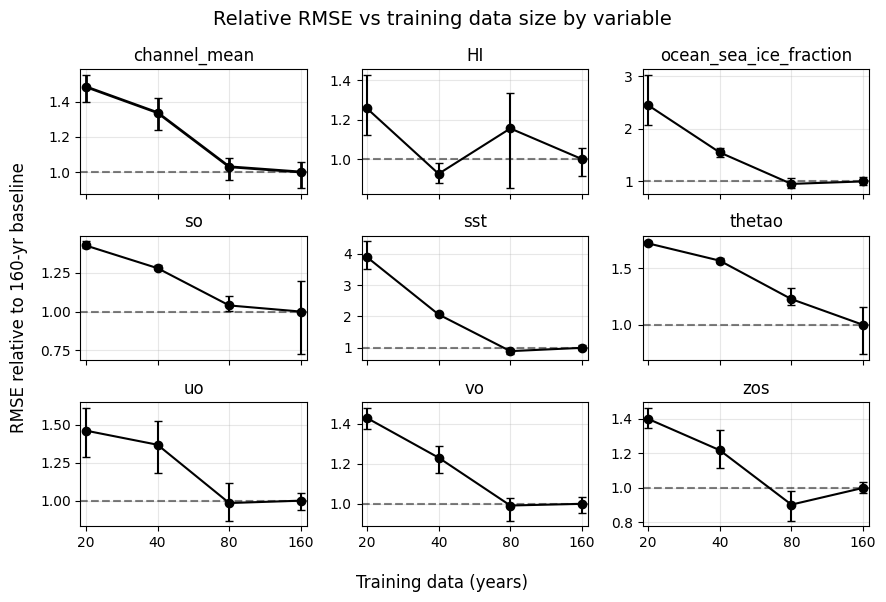}
\caption{Samudra's 40-year time and global mean normalized RMSEs vs training dataset size in years, with the mean, min, and max RMSE across three random initialization runs. RMSEs are normalized by the mean of the RMSEs achieved with 160 years of training data.}
\label{fig:rmse-vs-traindata}
\end{figure}

\begin{figure}
\centering\includegraphics[width=0.6\textwidth]{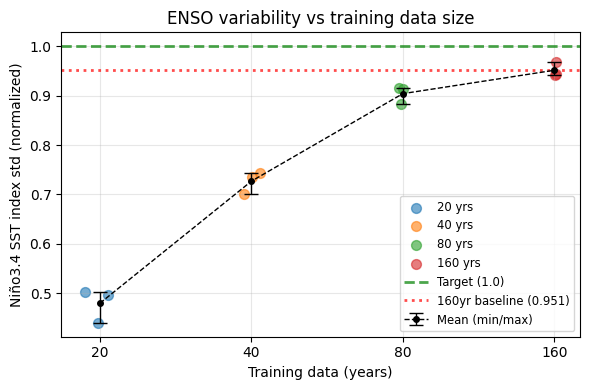}
\caption{Normalized standard deviation of the Niño 3.4 SST index vs training dataset size in years. The generated Niño 3.4 index standard deviation is normalized by the standard deviation of the CM4 reference index.}
\label{fig:enso-vs-traindata}
\end{figure}

\begin{figure}
\centering\includegraphics[width=0.8\textwidth]{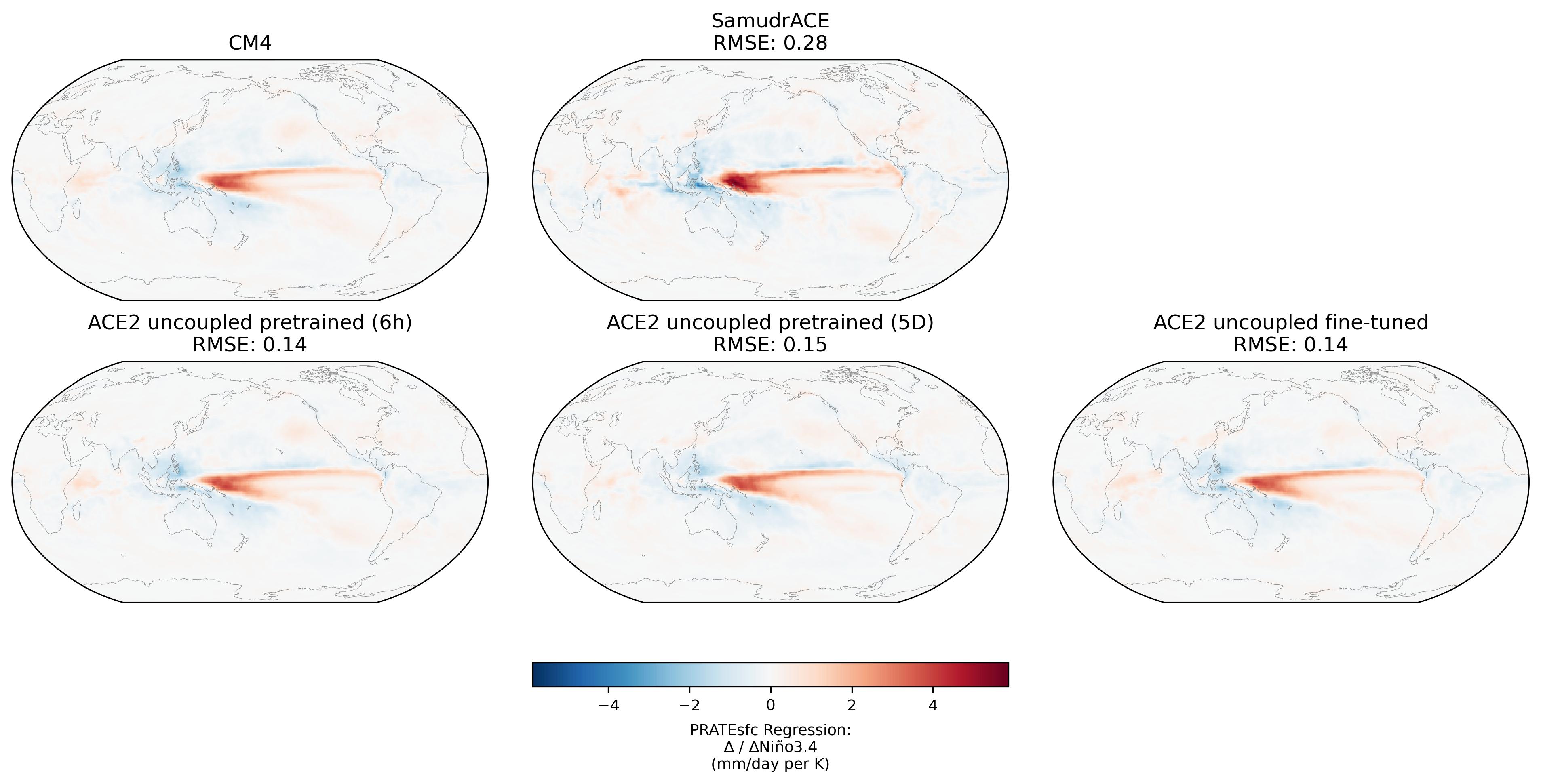}
\caption{Linear response of CM4 reference precipitation to the corresponding Niño 3.4 index is compared against the generated precipitation response. In all maps the CM4 reference Niño 3.4 index is used as the independent variable in the regression, with the exception of the SamudrACE case where SST is generated by Samudra and therefore the predicted Niño 3.4 index is used in the regression. The maps labeled ``CM4" and ``SamudrACE" are both shown in Figure 4 of the main text and repeated here for reference. The map labeled ``ACE2 uncoupled pretrained (5D)" shows the ACE2-generated precipitation response when forced by prescribed CM4 reference SSTs for the ACE uncoupled pretraining checkpoint that was used to initialize SamudrACE training. The map labeled ``ACE2 uncoupled fine-tuned" is the subset of SamudrACE's weights that comprise the ACE2 component of the coupled emulator. In particular, ``ACE2 uncoupled fine-tuned" is an ancestor of ``ACE2 uncoupled pretrained", separated by the coupled fine-tuning of the ``ACE2 uncoupled pretrained" weights that occurred while training SamudrACE. The ``fine-tuned" ACE2 precipitation response to the reference CM4 Niño 3.4 index has slightly lower RMSE than the ``pretrained (5D)" response. The map labeled ``ACE2 uncoupled pretrained (6h)" comes from another uncoupled ACE2 pretraining model that was trained with 6-hour varying SST.}
\label{fig:ace2_precip_response}
\end{figure}

\begin{figure}
\centering\includegraphics[width=0.7\textwidth]{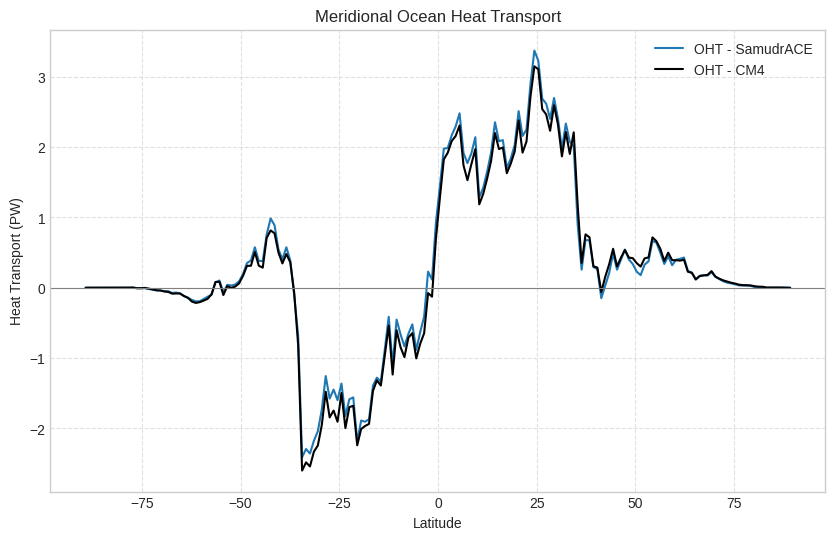}
\caption{Meridional ocean heat transport (North positive).}
\label{fig:oht}
\end{figure}

\begin{figure}
\centering\includegraphics[width=1\textwidth]{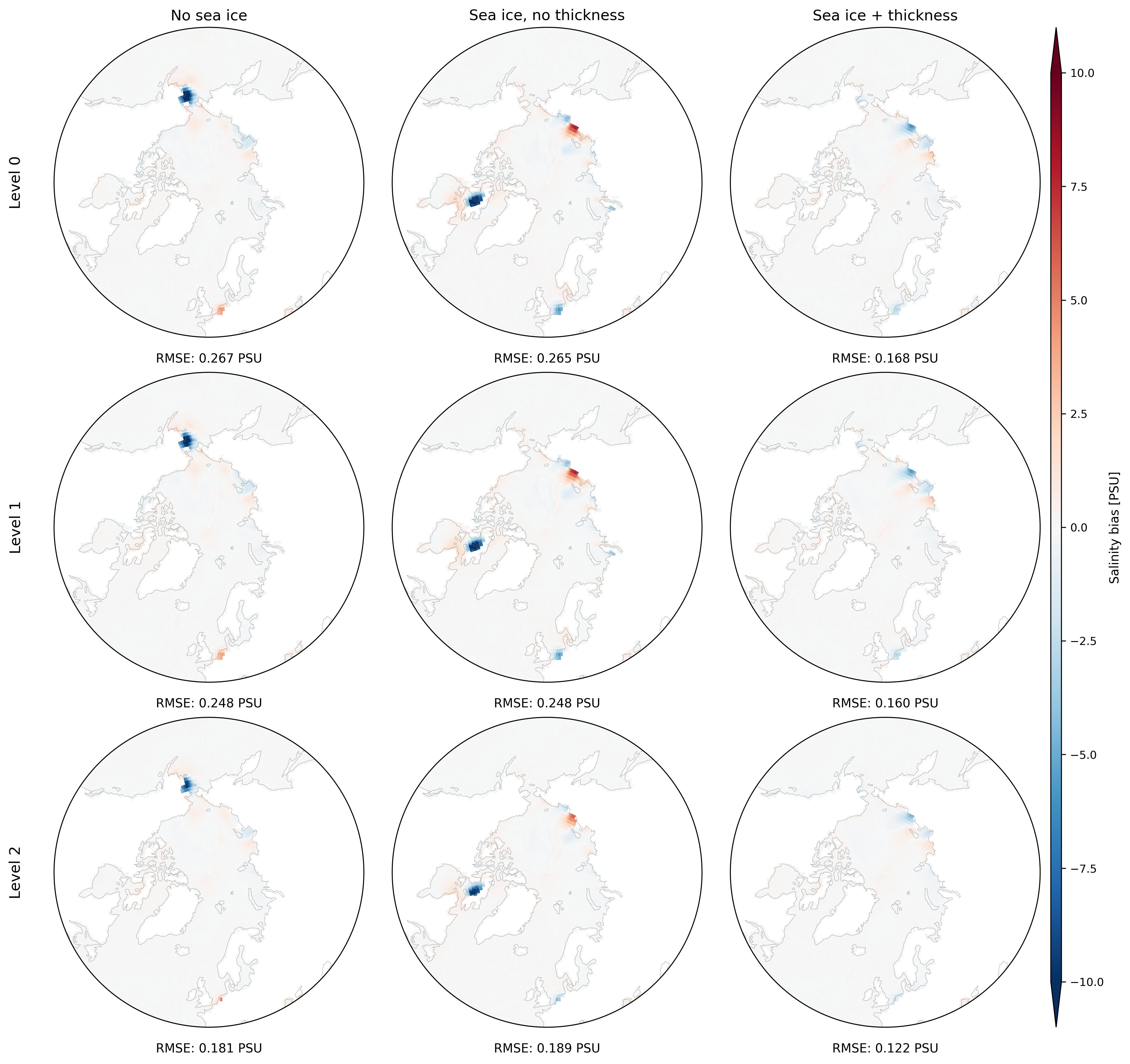}
\caption{North polar time mean bias maps and corresponding global RMSEs for the top 3 levels of salinity under various configurations of uncoupled Samudra. The first column shows salinity biases when Samudra is not trained to output sea ice fractions. The second column shows the same biases when Samudra is trained to predict the sea ice fraction but not sea ice thickness. The final column shows the biases when Samudra predicts both sea ice and thickness.}
\label{fig:effect-of-sea-ice}
\end{figure}

\begin{figure}
\centering\includegraphics[width=1\textwidth]{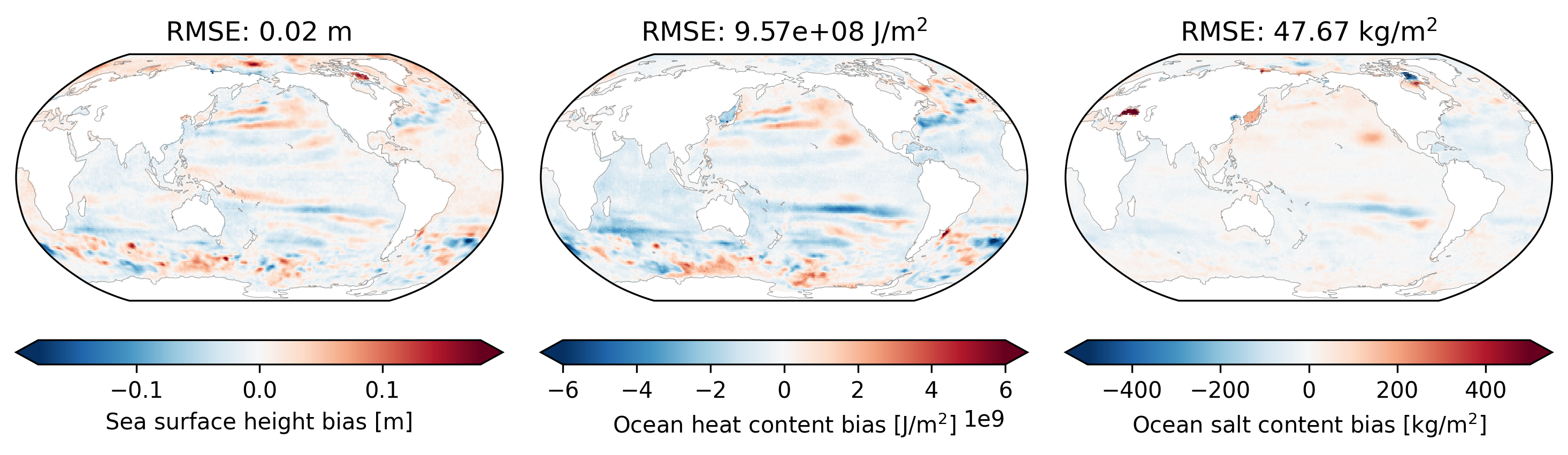}
\caption{Sea surface height, ocean heat content, and salt content bias over the 40 year test period.}
\label{fig:ssh-ohc-salt-bias-map}
\end{figure}

\begin{table}
  \caption{Description of SamudrACE sea ice and ocean variables shown in the left column of Figure \ref{fig:time_mean_rmse}. All variables are prognostic (input and output) and 5-day time averaged. Three dimensional variables have 18 levels, denoted by the subscript $k \in \{0 \ldots, 18\}$. Additional variables beyond what was included in \citeA{dheeshjith2025samudra} have descriptions highlighted in bold.}
  \label{table:variables_samudra}
  \centering
  \begin{tabular}{lll}
    \hline \\
    Symbol & Description & Units \\
    \hline \\
    $SIC$ & \textbf{Sea ice fraction of the ocean surface} & $[0, 1]$ \\
    $HI$ & \textbf{Sea ice thickness} & m \\
    $SST$ & Sea surface temperature & K \\
    $ZOS$ & Sea surface height above the geoid & m \\
    $\theta_{o,k}$ & Sea water velocity in the northward direction & °C \\
    $s_{o,k}$ & Sea water salinity & PSU \\
    $u_{o,k}$ & Sea water velocity in the eastward direction & m/s \\
    $v_{o,k}$ & Sea water velocity in the northward direction & m/s \\
    \hline
  \end{tabular}
\end{table}

\begin{table}
  \caption{
  Description of SamudrACE atmosphere output variables shown in the right column of Figure \ref{fig:time_mean_rmse}.
  Three dimensional variables have 8 levels, denoted by the subscript $k \in \{0 \ldots, 7\}$.
  }
  \label{table:variables_ace}
  \centering
  \begin{tabular}{lllll}
    \hline \\
    Symbol   & Description                                & Units & Time & Prognostic? \\
    \hline \\
    $T_k$    & Air temperature                            & K     & 6-hour snapshot & Yes \\
    $T_s$     & Skin temperature of land or sea-ice            & K     & Snapshot & Yes \\
    $q^T_k$   & Specific total water (vapor + condensates) & g/kg & Snapshot & Yes \\
    $U_k$    & Windspeed in eastward direction            & m/s   & Snapshot & Yes \\
    $V_k$    & Windspeed in northward direction           & m/s   & Snapshot & Yes \\
    $p_s$     & Atmospheric pressure at surface            & hPa    & Snapshot & Yes \\
    $RSW$ & Upward shortwave radiative flux at TOA         & W/m$^2$ & Mean & No \\
    $OLR$ & Upward longwave radiative flux at TOA          & W/m$^2$ & Mean & No \\
    $USW_{sfc}$ & Upward shortwave radiative flux at surface     & W/m$^2$ & Mean & No \\
    $ULW_{sfc}$ & Upward longwave radiative flux at surface      & W/m$^2$ & Mean & No \\
    $DSW_{sfc}$ & Downward shortwave radiative flux at surface   & W/m$^2$ & Mean & No \\
    $DLW_{sfc}$ & Downward longwave radiative flux at surface    & W/m$^2$ & Mean & No \\
    $LHF$ & Surface latent heat flux                            & W/m$^2$ & Mean & No \\
    $SHF$ & Surface sensible heat flux                          & W/m$^2$ & Mean & No \\
    $P$ & Surface precipitation rate (all phases)               & mm/day & Mean & No \\
    $\left. \frac{\partial TWP}{\partial t}\right |_{adv}$ & Tendency of total water path from advection & mm/day & Mean & No \\
    $\tau_u$ & Eastward surface wind stress & Pa & Mean & No \\
    $\tau_v$ & Northward surface wind stress & Pa & Mean & No \\~\\
    \hline
  \end{tabular}
\end{table}

\end{document}